\documentclass[aps,pre,superscriptaddress,twocolumn,balancelastpage,10pt]{revtex4-1}

\usepackage{amsmath}
\usepackage{amssymb}
\usepackage{hyperref}
\usepackage{graphicx}
  \graphicspath{{/}{resub-arxiv/}}
\usepackage{verbatim}
\usepackage{bbm}

\usepackage{color}
\definecolor{TextGreen}{rgb}{0,0.6,0}
\definecolor{TextRed}{rgb}{0.9,0,0}
\definecolor{TextBlue}{rgb}{0,0,0.9}

\newcommand{\eqn} {Eq.~}
\newcommand{\eqns}{Eqs.~}

\newcommand {\avg}[1]       {\langle#1\rangle}
\newcommand {\ee}           {\mathrm{e}}
\newcommand {\ii}           {\mathrm{i}}
\newcommand {\ket}[1]       {\lvert#1\rangle}
\newcommand {\bra}[1]       {\langle#1\rvert}
\newcommand {\braket}[2]    {\langle#1\vert#2\rangle}
\newcommand {\bramidket}[3] {\langle#1\vert#2\vert#3\rangle}
\newcommand {\abs}[1]       {\lvert#1\rvert}
\newcommand {\norm}[1]      {\lVert#1\rVert}
\newcommand {\spinup} {\mathnormal{\uparrow}}
\newcommand {\spinUp} {\mathnormal{\Uparrow}}
\newcommand {\spindn} {\mathnormal{\downarrow}}
\newcommand {\spinDn} {\mathnormal{\Downarrow}}
\newcommand {\trion}  {\mathrm{T}}

\newcommand {\half}   {\tfrac{1}{2}}
\newcommand {\Half}   {\frac{1}{2}}

\newcommand {\Li}     {\mathcal{L}}  
\newcommand {\ET}     {E_\trion}
\newcommand {\Lar}    {\Lambda}      
\newcommand {\lar}    {\lambda}      

\newcommand {\qf}     {\theta}
\newcommand {\Tpulse} {T_\mathrm{pulse}}
\newcommand {\Tcoh}   {T_\mathrm{coh}}
\newcommand {\Tillum} {T_\mathrm{illum}}
\renewcommand {\AA}   {\mathcal{A}}

\newcommand  {\muB}   {\mu_\mathrm{B}}

\newcommand  {\ps}    {\,\mathrm{ps}}
\newcommand  {\ns}    {\,\mathrm{ns}}
\newcommand  {\mus}   {\,\mu\mathrm{s}}
\newcommand  {\eV}    {\,\mathrm{eV}}
\newcommand  {\meV}   {\,\mathrm{meV}}
\newcommand  {\mueV}  {\,\mu\mathrm{eV}}
\newcommand  {\Tesla} {\,\mathrm{T}}

\newcommand  {\pplus}   {\bar{p}\mathnormal{+}}
\newcommand  {\pminus}  {\bar{p}\mathnormal{-}}

\newcommand  {\pplusq}  {\bar{p}\mathnormal{+}\;q}
\newcommand  {\pminusq} {\bar{p}\mathnormal{-}\;q}
\newcommand  {\pqplus}  {p\;\bar{q}\mathnormal{+}}
\newcommand  {\pqminus} {p\;\bar{q}\mathnormal{-}}

\newcommand {\Tr}    {\mathop{\mathrm{Tr}}\nolimits}
\newcommand {\diag}  {\mathop{\mathrm{diag}}\nolimits}
\renewcommand {\Re}  {\mathop{\mathrm{Re}}\nolimits}
\newcommand {\pe}[1] {^{(#1)}}   

\begin{document}
\title{Quantum model for mode locking in pulsed semiconductor quantum dots}
\date{\today}
\author{W. Beugeling}
\affiliation{Lehrstuhl f\"ur Theoretische Physik I, Technische Universit\"at Dortmund, Otto-Hahn-Stra\ss e 4, 44221 Dortmund, Germany}
\affiliation{Lehrstuhl f\"ur Theoretische Physik II, Technische Universit\"at Dortmund, Otto-Hahn-Stra\ss e 4, 44221 Dortmund, Germany}
\author{G\"otz S. Uhrig}
\affiliation{Lehrstuhl f\"ur Theoretische Physik I, Technische Universit\"at Dortmund, Otto-Hahn-Stra\ss e 4, 44221 Dortmund, Germany}
\author{Frithjof B. Anders}
\affiliation{Lehrstuhl f\"ur Theoretische Physik II, Technische Universit\"at Dortmund, Otto-Hahn-Stra\ss e 4, 44221 Dortmund, Germany}

\begin{abstract}

Quantum dots in GaAs/InGaAs structures have been proposed as a candidate system for realizing quantum computing. The short coherence time of the electronic quantum state that arises from coupling to the nuclei of the substrate is dramatically increased if the system is subjected to a magnetic field and to repeated optical pulsing. This enhancement is due to mode locking: Oscillation frequencies resonant with the pulsing frequencies are enhanced, while off-resonant oscillations eventually die out. Because the resonant frequencies are determined by the pulsing frequency only, the system becomes immune to frequency shifts caused by the nuclear coupling and by slight variations between individual quantum dots. The effects remain even after the optical pulsing is terminated. In this work, we explore the phenomenon of mode locking from a quantum mechanical perspective. We treat the dynamics using the central spin model, which includes coupling to $10$--$20$ nuclei and incoherent decay of the excited electronic state, in a perturbative framework. Using scaling arguments, we extrapolate our results to realistic system parameters. We estimate that the synchronization to the pulsing frequency needs time scales in the order of $1\,\mathrm{s}$.

\end{abstract}

\maketitle

\section{Introduction}

Ground-breaking insights in the field of quantum computing have demonstrated that there is a class of computational problems that can be solved much more efficiently on a quantum computer than on a classical one \cite{Feynman1982,Deutsch1985,*DeutschJozsa1992,DiVincenzo1995}. This fascinating prospect has inspired a large amount of research directed at finding reliable realizations of quantum computers \cite{*[{}][{ and references therein.}] KloeffelLoss2013}. An essential but challenging requirement for successfully implementing quantum algorithms is to maintain sufficiently long coherence times \cite{DiVincenzo1995,BurkardEA2000}. 

One promising approach utilizes electronic spin states in quantum dots in solid-state systems \cite{LossDiVincenzo1998,ImamogluEA1999}. Quantum dots have been realized in several forms in semiconductor materials, such as InGaAs \cite{KloeffelLoss2013}. The self-assembled variety can be engineered by strain between two different semiconductor materials (e.g., GaAs and InGaAs) with a slight lattice-constant mismatch. The manipulation and readout is done optically \cite{KikkawaAwschalom1998,*KikkawaAwschalom1999,ShabaevEA2003,Glazov2012PSS}, which poses an advantage compared to interaction via magnetic fields.

In a simplified picture, a self-assembled quantum dot can be described as a single electron whose spin dynamics is subject to a fixed external magnetic field and to a small Overhauser field, the effective magnetic field that arises from a hyperfine interaction with the nuclei of the substrate \cite{Overhauser1953,MerkulovEA2002,SchliemannEA2003}. Because the electron is delocalized non-uniformly, the electron couples differently to each of the nuclei \cite{MerkulovEA2002,ImamogluEA2003}. The electron-spin dynamics is dominated by Larmor precession with a frequency set by the external magnetic field plus a statistical deviation due to the Overhauser field. This mechanism causes dephasing of the Larmor oscillations on a nanosecond time scale \cite{MerkulovEA2002,SchliemannEA2003,DuttEA2005,BraunEA2005,HansonEA2007,UrbaszekEA2013}, severely limiting the coherence time at first sight.

However, optical excitation of the electron with periodically applied short (picosecond) laser pulses can increase the coherence time dramatically \cite{ImamogluEA2003,HuttelEA2004,BrackerEA2005,GreilichEA2006Science,GreilichEA2007,ReillyEA2008,*XuEA2009,*ChekhovichEA2010,*SunEA2012,*TenbergEA2015}. The effect of the (pump) pulses is two-fold: They quickly drive the system towards a steady state, which exhibits a revival effect of the dephased Larmor oscillations \cite{GreilichEA2007}. Secondly, the full system slowly becomes synchronized to the pulsing repetition rate \cite{GreilichEA2006Science,LattaEA2009,LaddEA2010,YugovaEA2009,YugovaEA2012,GlazovEA2012PRB,EconomouBarnes2014}. Any mode that is not resonant with the pulse repetition rate eventually dies out. This effect, known as mode locking, manifests itself as an additional enhancement of the amplitude of the revivals. In practice, advanced schemes composed of multiple pump pulses per period tend to increase coherence times even further \cite{GreilichEA2007,VarwigEA2016}.

The typical experimental scenario is an ensemble of quantum dots rather than a single one \cite{YugovaEA2009}. The individual quantum dots differ slightly, e.g., in their effective $g$ factors and hyperfine coupling strengths. As a consequence, the characteristic frequencies of the dots vary, but the resonant frequencies remain pinned at fixed values set by the pulsing period only. Thus, the revival effect is robust against these variations, and can indeed be observed in quantum dot ensembles \cite{GreilichEA2006PRL,GreilichEA2006Science,SpatzekEA2011PRB}.

A significant difference between two types of revivals is observed when the pulsing is terminated at some moment. Without mode locking, revivals appear after the pulsing ends, but they quickly attenuate. However, a quantum dot ensemble that has become mode locked after an extended exposure to periodic pulsing, will show strong revivals for a longer period \cite{GreilichEA2006Science} and thus retain coherence after the pulsing has been switched off. Mode locking is thus considered as the main mechanism responsible for the observed long coherence times, and consequently as an essential ingredient that renders pulsed quantum dots suitable for quantum-computational applications.

In this work, we aim for a theoretical explanation of mode locking by analysis of a minimal model for the spin dynamics in a single quantum dot. For this purpose, we use an extended version of the central spin model, also known as the Gaudin model \cite{Gaudin1976}. This integrable \cite{FaribaultSchuricht2013PRL,*FaribaultSchuricht2013PRB} model incorporates the external magnetic field and the hyperfine couplings between the electron and nuclear spins \cite{KhaetskiiEA2002,SchliemannEA2003}. We additionally include the optical interaction as instantaneous excitation of the electron to an excited (trion) state, which decays gradually. Because the latter process is incoherent, the system evolves non-unitarily. Here, we describe this time evolution effectively with the Lindblad formalism \cite{BreuerPetruccione2007book}.

Despite the apparent simplicity of the model, the theoretical description of the dynamics is challenging due to the vastly different time scales associated to the various interactions: the duration of each laser pulse takes place on a picosecond scale, and their repetition rate is typically $13.2\ns$ \cite{GreilichEA2006PRL,GreilichEA2006Science}. On the other hand, coherence can be maintained in time intervals spanning several minutes or longer \cite{HuttelEA2004,GreilichEA2007}. In addition, the number of degrees of freedom in the system grows exponentially with the number $N$ of nuclei in the model. Thus, a full quantum mechanical description is feasible only for very small numbers of nuclei, typically $N\sim 10$, which is far from the realistic value of $N\sim10^4$--$10^6$ \cite{MerkulovEA2002,SchliemannEA2003,ImamogluEA2003}.

We tackle this problem with a perturbative approach for the time evolution. The underlying idea is the separation of time scales between the Larmor precession and the trion decay on one hand, and the hyperfine interaction on the other. The hyperfine coupling is separated into a longitudinal part (parallel to the external field) and a transverse part. The latter is treated as a perturbation to lowest non-trivial order. We justify this approach from the time scale of higher-order perturbations being much longer than the pulse interval. One key advantage of this method is that we obtain analytic estimates for the resonant frequencies. We also use the perturbed results for numerics at long time scales, up to $\sim 20000$ pulse intervals, which corresponds to $\sim 200 \mus$.

Our numerical method is not capable of reaching experimentally relevant regimes in terms of system size and times, but from our results we obtain scaling laws that allow us to extrapolate. We study the relative difference between the spectrum after a long period of pulsing and the initial one. In doing so, we find tiny but robust peaks at the frequency values where we expect the resonances to be. The growth rate of these peaks turns out to be quadratic in the hyperfine coupling strength. We also  investigate the dependence on the modeled number of nuclei $N$ and the effect of the discretization of the distribution of coupling strengths. Our eventual estimate for the required pulsing duration is of the order of $\sim0.1$--$1\,\mathrm{s}$, consistent with experimental observations \cite{GreilichBelykh_privatecommunication}.

In this article we proceed as follows. In Sec.~\ref{secModel}, we set up our model. Section~\ref{secPert} is dedicated to the time evolution in a general sense and to the perturbative framework. We provide and interpret the results on the mode-locking effect in Sec.~\ref{secModeLocking}. We conclude in Sec.~\ref{secConclusion} with a discussion and an outlook. In the Appendices, we provide technical details on the perturbative method and a steady-state analysis.

\section{Model}
\label{secModel}
Our aim is to describe the dynamics of the central spin and the nuclear spins in the quantum dot that mutually interact through the hyperfine coupling, and are subject to an external magnetic field and to laser pulses that excite the central-spin electron to the trion state. We consider the system in a Voigt geometry, where the magnetic axis ($\parallel \hat{x}$) and the optical axis ($\parallel \hat{z}$) are perpendicular.

The degrees of freedom associated to the central spin are given by four basis states: two ground states $\ket{\spinup}$ and $\ket{\spindn}$ and two excited (trion) states $\ket{\spinUp\spinup\spindn}$ and $\ket{\spinDn\spinup\spindn}$. Typically, the laser radiation is circularly polarized in one single helicity \cite{ShabaevEA2003,SpatzekEA2011PRL}, so that one of the trion states decouples. We therefore restrict ourselves to a three-dimensional Hilbert space for the central spin, with the basis $\{\ket{\spinup}, \ket{\spindn},\ket{\trion}\}$ where $\ket{\trion}\equiv\ket{\spinUp\spinup\spindn}$ encodes the trion state that is relevant to the dynamics.

We customarily treat the nuclei as effective spin-$\half$ particles, although in fact, the nuclei in question have higher spin quantum numbers of $\tfrac{3}{2}$ (for Ga and As) or $\tfrac{9}{2}$ (for In). Within the scope of this work, where the only nuclear interaction is the hyperfine coupling to the central spin, this simplification does not lead to essentially different physics. Thus, given $N$ spin-$\half$ nuclei in addition to the central spin, we have a total Hilbert space dimension of $D=3 \times 2^N$.

The coherent part of the dynamics in the central spin model is described by a Hamiltonian that encodes the effect of the external magnetic field and the hyperfine coupling between the central spin and the nuclear spins \cite{KhaetskiiEA2002,SchliemannEA2003,UrbaszekEA2013},
\begin{equation}\label{eqnHamiltonian}
  H = \Lar \hat{S}^x + \ET \ket{T}\bra{T}+ \sum_{j=1}^N A_j (\hat{I}^x_j\hat{S}^x + \hat{I}^y_j\hat{S}^y + \hat{I}^z_j\hat{S}^z).
\end{equation}
where $\hat{S}^\mu$ and $\hat{I}^\mu_j$ ($\mu=x,y,z$) are the components of the spin operators of the central spin and the nuclear spins, respectively, in units of $\hbar$. The first term encodes the Larmor precession due to the external magnetic field. We denote the associated energy by $\Lar =g \muB B_\mathrm{ext}$ in terms of the Land\'e $g$ factor, the Bohr magneton $\muB$ and the external magnetic field $B_\mathrm{ext}$ ($\vec{B}_\mathrm{ext}=B_\mathrm{ext}\hat{x}$). The second term sets the trion state at an energy $\ET$ relative to the central spin states $\ket{\spinup}$ and $\ket{\spindn}$. The third term is the hyperfine coupling between the central spin and each of the nuclear spins. The coupling strengths are encoded through the energies $A_j$. For the sake of simplicity, we neglect the effect of the external magnetic field on the nuclear spins, and omit any additional couplings that are relevant only at time scales much longer than the pulse repetition period, such as the quadrupolar coupling term between the electron and the nuclei \cite{SinitsynEA2012,HackmannEA2015,BechtoldEA2015,SokolovEA2016} or the hyperfine interaction among the nuclear spins \cite{AuerEA2009}.

Here, we notice the vastly different energy scales in this Hamiltonian. The trion energy $\ET$ typically has a value of $1.39\eV$ \cite{GreilichEA2006PRL}. The Larmor energy $\Lar$ lies in the range of a few $0.1\meV$ for typical fields of $6\Tesla$. The values of the couplings $A_j$ depend on the details of the system, e.g., the localization area of the central-spin electron in the sample. Generally, they are much smaller than $\Lar$ for the range of external fields we consider. Typical values for the largest couplings lie in the $\mu\mathrm{eV}$ range. The corresponding time scales for the Larmor and hyperfine oscillations are $20\ps$ and $10^3$--$10^4\ps$, respectively \cite{MerkulovEA2002}.

The relevant time scales of the Hamiltonian dynamics are determined by the frequencies $\lambda\equiv\Lambda/\hbar$ and $a_j\equiv A_j/\hbar$. When $a_j/\lar$ is small, as we shall assume throughout this work, the effect of the hyperfine coupling is to shift the eigenfrequencies slightly away from the Larmor frequency $\lar$. This effect is observable as dephasing of the Larmor precession \cite{MerkulovEA2002}. The characteristic time scale, known as the dephasing time $T^*$, is determined by the squared sum of the couplings $\AA=\sum_{j=1}^{N}a_j^2$ as $T^* = \sqrt{8 / \AA}$. By virtue of the central limit theorem, for a fixed value of $\AA$, the choice of the individual couplings $a_j$ does not affect the dephasing essentially\cite{MerkulovEA2002,Glazov2012PSS}.

The latter statement is not \emph{a priori} true for higher-order effects, such as the synchronization to the pulses that we shall focus on. Let us therefore specify a realistic choice of the values $a_j$ of hyperfine couplings, based on the idea that $a_j$ is proportional to the probability density of the electronic wave function at the position of the nucleus labeled by $j$. In good approximation, we assume that the wave function amplitude is a Gaussian in two dimensions $\propto\ee^{-\abs{r}^2/2R^2}$ which is cut off at $r=r_\mathrm{cutoff}$, based on the idea that the electron is confined to a finite region. Assuming that each nucleus occupies an equal area, we set
\begin{equation}\label{eqnCouplings}
  a_j = \frac{C}{2\pi R^2}\exp\left(-\frac{j}{N+1} \frac{r_\mathrm{cutoff}^2}{2R^2}\right)
  \qquad (j=1,\ldots,N),
\end{equation}
where $R$ is a scaling factor denoting the characteristic radius of the electronic wave function, cf. Refs.~\cite{CoishLoss2004,BortzEA2010,FaribaultSchuricht2013PRL,*FaribaultSchuricht2013PRB,VanDenBergEA2014,HackmannAnders2014,SeifertEA2016,GravertEA2016}. The constant $C$ is set such that the square sum $\sum_j a_j^2$ has the value corresponding to the dephasing time $T^*$, which we treat as an input parameter.

The shape of the distribution is determined only by the dimensionless cutoff parameter $\tilde{r}_\mathrm{cutoff} =r_\mathrm{cutoff}/R$. For small values ($\tilde{r}_\mathrm{cutoff} \to 0$), all couplings are (almost) equal, which is the so-called \emph{box model}, named after the idea that the wave function amplitude can be thought of as constant. For large values of $\tilde{r}_\mathrm{cutoff}$, the distribution contains a few larger couplings and relatively many small ones, where the latter correspond to weakly interacting nuclei in the tail of the Gaussian wave function. This may be understood from the observation that in the continuous limit $N\to\infty$, the probability density function $D(a)$ corresponding to \eqn\eqref{eqnCouplings} is proportional to $1/a$ with appropriate cutoffs, with the lower one set by $\tilde{r}_\mathrm{cutoff}$. For large $\tilde{r}_\mathrm{cutoff}$, the lower cutoff of $D(a)$ is small, and the distribution is then dominated by small values of $a$, i.e., the weak couplings. In the remainder of this work, we choose  the value $\tilde{r}_\mathrm{cutoff}=2$, unless stated otherwise.

\section{Time evolution}
\label{secPert}
\subsection{General framework}
\label{subsecEvolutionGeneral}

The full unitary time evolution of the system including the excitation and decay of the trion would require that we include the photons it absorbs and emits as part of the Hilbert space. This full problem being intractable, we treat the photon degrees of freedom effectively through the Lindblad formalism \cite{BreuerPetruccione2007book}. In this formalism, the spin degrees of freedom constitute our ``system'', whereas the photons are treated as the ``bath''. This effective description comes at the cost of losing unitarity in the dynamics of the system part. Physically speaking, the 
trion decay acts incoherently on the system. Energy is not necessarily conserved: the trion decay is due to photons which carry energy from the system into the photonic bath. In this formalism, the system is described by a density matrix $\rho(t)$ rather than by a quantum state. The Lindblad master equation that governs the dynamics of the density matrix is
\begin{equation}\label{eqnLindblad}
  \frac{d\rho}{dt}(t) = \Li \rho(t)
\end{equation}
where $\Li$ is the Liouville operator that acts as
\begin{equation}\label{eqnLiouville}
\Li \rho = -\frac{\ii}{\hbar} [H,\rho] - \gamma \left( \half b^\dagger b \rho + \half \rho b^\dagger b - b\rho b^\dagger\right).
\end{equation}
The first term describes the unitary part of the dynamics, involving the Hamiltonian of \eqn\eqref{eqnHamiltonian}. The second term constitutes the single decoherence channel of the trion decay, with operator $b=\ket{\spinup}\bra{\trion}$ acting on the central spin only. The decay rate $\gamma$ is typically of the order of $(400\ps)^{-1}$ \cite{GreilichEA2006PRL}; the energy equivalent is $\hbar\gamma\sim 1\mueV$.

In comparison to the Larmor oscillations, the hyperfine interaction, and the trion decay rate, the duration of the pulses (up to $1\ps$) is sufficiently short that they can effectively be considered as instantaneous: Each moment the system is pulsed, the state of the central spin is unitarily mapped $\ket{\psi}\mapsto \mathcal{P}\ket{\psi}$ \cite{YugovaEA2012,BarnesEconomou2011,EconomouBarnes2014}; in density-matrix language, the pulse action reads as $\rho\mapsto \mathcal{P} \rho \mathcal{P}^\dagger$. In this work, we consider $\pi$-pulses only, which map the central-spin state $\ket{\spinup}$ to the trion state $\ket{\trion}$, and leaves $\ket{\spindn}$ invariant. The corresponding pulse action thus reads $\mathcal{P}=\ket{\trion}\bra{\spinup} - \ket{\spinup}\bra{\trion} + \ket{\spindn}\bra{\spindn}$.

Because the Liouville operator is time-independent, the Lindblad equation can be solved formally as
\begin{equation}\label{eqnLindbladSolution}
 \rho(t) = \ee^{t\Li} \rho(0).
\end{equation}
Whereas the solution is formally simple, a concrete solution involves diagonalization of the Liouville operator $\Li$ in order to compute the exponential. The Liouville operator is a linear operator on the $D^2$ dimensional vector space of density matrices; the expression $\Li\rho$ (\eqn\eqref{eqnLiouville}) does not represent a matrix multiplication of two $D\times D$ matrices, but should be interpreted as a matrix multiplication of a $D^2\times D^2$ matrix and a $D^2$ component vector. A brute-force calculation of \eqn\eqref{eqnLindbladSolution} would thus require diagonalization of a matrix of dimension $D^2\times D^2$. As the Hilbert-space dimension $D$ grows exponentially in the number of nuclei $N$, the brute-force approach becomes intractable for anything more than a few spins. This problem motivates the need for other methods of calculation.

\subsection{Perturbation theory}

The key idea behind the perturbative treatment is the separation of time scales. We consider the Larmor precession and the trion decay as ``fast'' processes, and the hyperfine dynamics as ``slow''. In terms of the energy scales, the hyperfine couplings are much smaller than the other energies, namely $A_j\ll\Lar$ and $A_j\ll\hbar\gamma$. We thus include the fast dynamics in the zeroth order of the perturbation theory and treat the hyperfine dynamics perturbatively.

Because the aim is to obtain the dynamics governed by the Lindblad master equation, the object that is treated perturbatively is the Liouville operator $\Li$. Following the idea of separating the fast and slow dynamics, one would be tempted to choose $\Li\pe0$ as given by \eqn\eqref{eqnLiouville} with $H$ replaced by $H\pe0=\Lar\hat{S}^x + \ET\ket{\trion}\bra{\trion}$. The diagonalization of this Liouvillian is straightforward, but one runs into a high degree of degeneracy, because $\Li\pe0$ acts non-trivially only in the central-spin space and as the identity in the nuclear spin sector. Instead of dealing with the difficulties of highly degenerate perturbation theory, we include the $\hat{x}$ part of the hyperfine coupling into the zeroth order. We thus define $\Li\pe0$ and $\Li\pe1$ according to
\begin{subequations}
\begin{align}
\Li\pe0 \rho &= -\frac{\ii}{\hbar} [H\pe0,\rho] + \gamma \left(\half b^\dagger b \rho + \half\rho b^\dagger b - b\rho b^\dagger\right),%
\label{eqnLi0}\\
\Li\pe1 \rho &= -\frac{\ii}{\hbar} [H\pe1,\rho],%
\label{eqnLi1}
\end{align}
\end{subequations}
with
\begin{subequations}
\begin{align}
  H\pe0 &= \Lar \hat{S}^x + \ET \ket{T}\bra{T}+ \sum_{j=1}^N A_j \hat{I}^x_j\hat{S}^x,%
  \label{eqnH0}\\
  H\pe1 &= \sum_{j=1}^N \tilde{A}_j (\hat{I}^y_j\hat{S}^y + \hat{I}^z_j\hat{S}^z)\nonumber\\
  &= \Half\sum_{j=1}^N \tilde{A}_j (I^+_jS^- + I^-_jS^+),%
  \label{eqnH1}
\end{align}
\end{subequations}
cf.\ Refs.~\cite{KhaetskiiEA2002,CoishEA2010}. Here, we used a different notation $\tilde{A}_j$ for the transverse couplings, with the same values $A_j$ in order to keep track of the perturbation parameters $\tilde{A}_j/\Lar$. The operators $S^\pm=S^z\mp\ii S^y$ and $I_j^\pm=I_j^z\mp\ii I_j^y$ are the raising and lowering operators in the spin-$\hat{x}$ basis for the central and nuclear spins, respectively. The Hamiltonian $H\pe1$ describes processes where spin $\hat{S}^x$ is transferred from the central electron to a nucleus and vice versa. 

The zeroth order time evolution, that involves diagonalization of $\Li\pe0$, is particularly straightforward in the basis of eigenstates of $S^x$ and $I^x_j$. In this basis, $\Li\pe0$ is diagonal, and we directly read off the eigenvalues $\pm\ii\omega_{pq}$, $\pm\ii\Omega_{pq}$, and $-\gamma$, where
\begin{align}
  \omega_{pq} &= \half(\qf^p-\qf^q),\nonumber\\
  \Omega_{pq} &= \lar + \half(\qf^p + \qf^q),\label{eqnOmegas}
\end{align}
with the definition
\begin{equation}\label{eqnQF}
  \qf^p\equiv \bramidket{p}{\sum_{j=1}^N a_j I^x_j}{p}
            = \sum_{j=1}^N a_js^p_j,
\end{equation}
which encodes frequency shifts of the central-spin oscillations induced by the hyperfine interaction with the nuclear spins in the $x$ direction. The indices $p$ and $q$ label configurations of the nuclear spin, i.e., states of the form $\ket{s^p_1,s^p_2,\ldots,s^p_{N}}$, where $s^p_j=\bramidket{p}{I^x_j}{p}$.

In this zero-order model, the presence of dephasing follows naturally from inclusion of the longitudinal component of the hyperfine interaction into the Hamiltonian $H\pe0$. The Larmor precession is represented by the expectation values $\avg{S^y}(t)$ and $\avg{S^z}(t)$. [For a generic observable $\hat{O}$, the time-dependent expectation value is given by $\avg{O}(t) = \Tr [\hat{O}\rho(t)]$.] As shown in Appendix~\ref{appFullPert}, $\avg{S^y}(t)$ and $\avg{S^z}(t)$ contain oscillatory contributions with the shifted Larmor frequencies $\Omega_{pp}=\lar+\qf^p$ [see \eqn\eqref{eqnOmegas}], in addition to decaying contributions (those involving exponentials of the form $\ee^{zt}$ with $\Re z<0$). Hence, the non-decaying contributions are a Fourier sum of the form
\begin{equation}\label{eqnFouriersum}
 \sum_{p} c_{pp}\ee^{\ii(\lar+\qf^p)t} + \mathrm{h.c.},
\end{equation}
where $p$ runs over all nuclear configurations and the coefficients $c_{pp}$ depend on the observable and on the initial density matrix. In the limit $a_j\ll\lar$ that we have assumed, the frequencies $\lar+\qf^p$ in \eqn\eqref{eqnFouriersum} all lie close to the bare Larmor frequency $\lar$. Even without exact details on the distribution of the couplings $a_j$, the central limit theorem implies that $\qf^p$ [\eqn\eqref{eqnQF}] has a distribution that is approximately Gaussian, with variance $\sigma^2=\frac{1}{4}\sum_ja_j^2=\frac{1}{4}\AA$. The latter quantity has a fixed value determined by the atomic properties of the quantum dot and by the amount of localization of the electronic wave function \cite{MerkulovEA2002}. If the coefficients are assumed to have equal weights ($c_{pp}\equiv c$), then the Fourier sum of \eqn\eqref{eqnFouriersum} is well approximated by the Fourier integral
\begin{equation}\label{eqnFourierintegral}
 \int\mathrm{d}\qf D(\qf)(c \ee^{\ii\qf t}\ee^{\ii\lar t} + \mathrm{h.c.})\\
  =\ee^{-t^2/2\sigma^2}(c \ee^{\ii\lar t} + c^*\ee^{-\ii\lar t})
\end{equation}
where $D(\qf)=\ee^{-\qf^2/2\sigma^2}/\sqrt{2\pi\sigma^2}$ is the normal distribution of the frequency shifts $\qf^p$. The right-hand side shows an oscillation with the Larmor frequency $\lar$ modulated by a Gaussian decay with characteristic time $1/\sigma\sqrt{2}=\sqrt{8/\AA}$. Here we observe the mechanism of dephasing: the contributions of slightly different frequencies gradually get out of phase, leading to a complete suppression of the oscillations at long times; see, e.g., Ref.~\cite{StanekEA2014} for an illustration. The characteristic time is the dephasing time $T^*$.

It should be noted that for the derivation of \eqn\eqref{eqnFourierintegral}, we have assumed a continuum limit, or equivalently, $N\to\infty$. For a finite and small number of nuclear spins, the dephasing is not perfect, and revivals occur, where oscillations accidentally ``re-phase'' at some time $t>0$. The typical time at which accidental revivals occur grows rapidly as function of $N$ however, and their amplitudes are negligible even for system sizes that we are able to treat numerically ($N\sim 15$--$20$), let alone for realistic values of $N\sim 10^4$--$10^6$.

The zeroth order captures the dephasing of the central spin due to the influence of the nuclear magnetic moments, but not the reverse effect of the central spin magnetic moment onto the nuclei. Mode locking cannot be described in this framework, because the nuclear dynamics does not respond to the pulsing directly, but only through coupling with the central spin. Non-trivial perturbations incorporate the nuclear spin flips essential for the nuclear dynamics that gives rise to mode locking.

The perturbation $\Li\pe1$ brings forth corrections to the eigenfrequencies [\eqn\eqref{eqnOmegas}] and to the eigenvectors. The first-order corrections to the eigenvalues all vanish, because a spin flip maps one nuclear configuration to another perpendicular one. The first-order correction to the eigenvectors, however, is highly non-trivial, and contains many terms that encode a single simultaneous flip of the central spin and one nuclear spin, see Appendix~\ref{appFullPert} for details.

The question arises as to whether expansion to first order for the eigenvalues and for the eigenvectors provides an accurate description that represents all essential aspects of the dynamics. In order to answer this question, we apply the perturbation theory to a minimal model, namely, the Hamiltonian dynamics of the central spin with a single nuclear spin. In Appendix~\ref{appMinimalmodel}, we compare the exact and perturbative time evolution in order to provide an estimate on how the errors scale in terms of the perturbation parameters $A_j$. The results suggest that for the eigenvalues, a second order perturbation is required, whereas for the eigenvectors, linear order is sufficient. Inclusion of higher orders order would increase the computational complexity by a considerable amount, while not improving the accuracy significantly.

\subsection{Numerical implementation}
\label{subsecNumerics}
The large Hilbert-space dimension poses a serious challenge for the numerical evaluation of the time evolution of the density matrix, even for the perturbative method. In order to be able to perform the calculation for moderate numbers of nuclei ($N\sim 15$--$20$), we store the density matrix in a sparse format, and compute the time evolution ``on-the-fly'' using the results exhibited in Appendix~\ref{appFullPert}. We do not store the Liouville operator explicitly, because it is generally too large even in a sparse format. The time evolved density matrix is again sparse, but with a larger number of nonzero entries: The number of nonzero entries is multiplied by up to $2N$ for each application of the first-order evolution operator, because the latter involves a spin flip at every nuclear spin, in either the row or the column index. Eventually, repeated application would lead to a dense (or an almost dense) matrix.

In order to limit the number of nonzero entries, we ``truncate'' the density matrix by neglecting all matrix entries whose magnitude is smaller than the predefined threshold value $4^{-(N+1)}$. Diagonal entries are exempt from truncation, in order to preserve the trace of the density matrix. Off-diagonal entries are generally small, as demonstrated by the structure of the perturbation theory, where each spin flip is accompanied by a small multiplication factor of approximately $a_j/\lar$. Furthermore, the decay and dephasing processes will additionally lead to exponential or Gaussian decay of some entries to values below the threshold. Thus, the threshold value can be kept quite low, so that the errors introduced by the truncation remain small. We justify this approximation with quantitative arguments involving the structure and size of the density matrix elements, presented in Appendix~\ref{appTruncation}.

The required computational resources scale exponentially in $N$. For the data presented in this work, we have restricted ourselves to $N\leq 17$. We consider the values $N=15$--$17$ as good compromise, for which the relevant physics is visible, at manageable computation times, typically up to a few $100$ CPU hours. Such computation times enable us to run multiple simultaneous computations for investigation of the dependence on external parameters, such as the dephasing time and the cutoff of the coupling distribution.

The initial density matrix is chosen to describe a completely disordered spin bath corresponding to a temperature scale that is essentially infinite from the perspective of the small energy scales in the Hamiltonian. Thus, the distribution of frequency shifts $\qf_p$ has a Gaussian shape centered at zero. The central spin is initially in the negative $z$ direction.
Generically, the initial configuration does not affect the results on long time scales. As we argue in Appendix~\ref{appSteadyState}, the system converges to a (quasi)steady state within a few pulse intervals, which is independent on the initial state. Mode locking is essentially a perturbation to this quasisteady state, and is thus unaffected by the initial configuration.

\section{Dynamics of the Overhauser field: Mode locking}
\label{secModeLocking}

\subsection{Overhauser spectrum} 
\label{subsecOverhauserSpectrum}

\begin{figure*}
\includegraphics{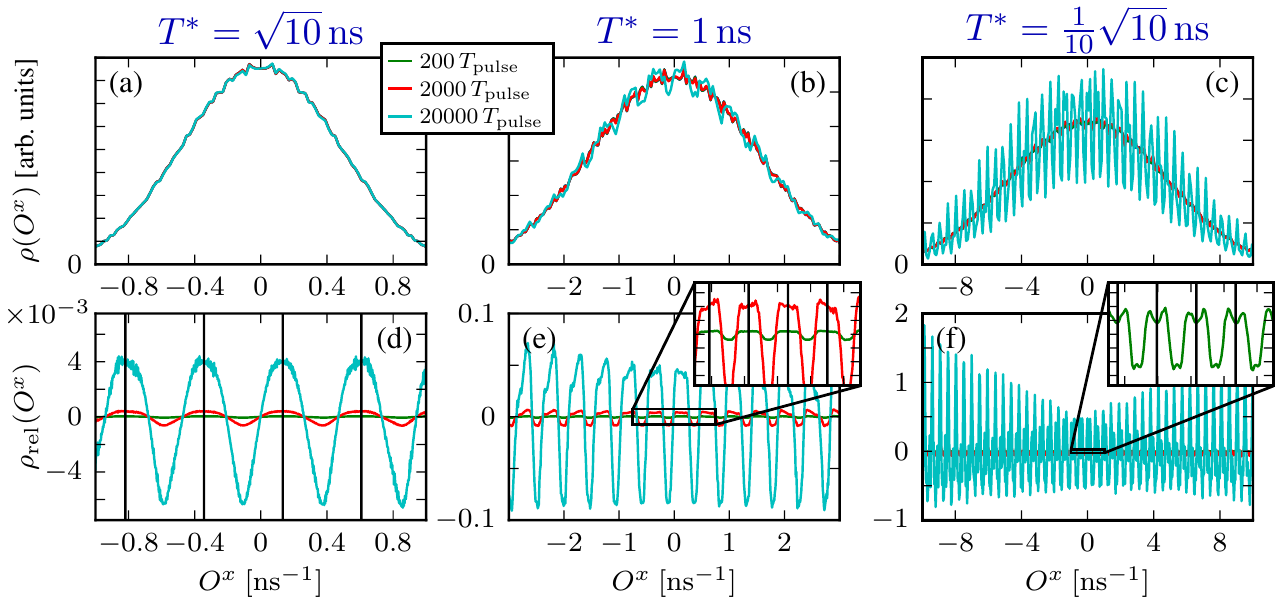}
\caption{(a) Spectrum of the longitudinal Overhauser field $O^x=\sum_ja_jI^x_j$ after $200$, $2000$ and $20000$ pulses, with realistic couplings $a_j$, chosen such that $T^*=\sqrt{10}\ns\approx3.16\ns$. (b, c) A similar plot for couplings multiplied by factors $\sqrt{10}$ and $10$, respectively, i.e., with $T^*=1\ns$ and $T^*=\tfrac{1}{10}\sqrt{10}\ns\approx0.316\ns$. In (a)--(c), the distributions are normalized to an integral of $1$. (d--f) Relative probability distributions $\rho_\mathrm{rel}(O^x)=\rho(O^x)_t/\rho(O^x)_0$. In (d) and in the insets of (e) and (f), the black vertical lines indicate the values where the resonance condition is fulfilled, according to \eqns\eqref{eqnResonanceCondition2} and \eqref{eqnTrionPhase}. The insets of (e) and (f) span the same horizontal and vertical range as panel (d).}
\label{figOverhauserSpectrum} 
\end{figure*}

In order to compare with other theoretical and experimental studies, we study mode locking through the Overhauser field $\vec{B}_\mathrm{O}$, the effective magnetic field caused by the nuclear spins. In particular, the longitudinal part (parallel to the external magnetic field) $B_\mathrm{O}^x$ shows strong signs of the mode locking effect, due to its \emph{almost} one-to-one correspondence with the oscillation frequencies. The latter frequencies are essentially the Larmor modes shifted by a contribution from the Overhauser field. Details on this correspondence will be given in Sec.~\ref{subsecResonance}.

In the following, we consider the observable $O^x=\sum_j a_j I^x_j = g\muB B^x_\mathrm{O}/\hbar$. Although $O^x$ has dimensions of (angular) frequency, we will refer to it as the ``Overhauser field'' as well, as it is proportional to the proper Overhauser field by the (dimensionful) constant $g\muB/\hbar$. The time dependent expectation value of $O^x$ reads as
\begin{equation}\label{eqnExpectationValue}
  \avg{O^x}(t)=\Tr [\rho(t)O^x] = \sum_p \rho_{pp}(t) O^x_{pp} = \sum_p \rho_{pp}(t) \qf_p
\end{equation}
where we have used the spin-$x$ basis, like in Sec.~\ref{secPert}. In this basis, $O^x$ is diagonal, $O^x_{pq}=\bramidket{p}{\hat{O}^x}{q}=\qf_p\delta_{pq}$. [The matrix element $\rho_{pp}(t)$ contains an implicit trace over the central-spin degrees of freedom.] In order to extract more information than just the expectation value, we interpret the summation $\sum_p \rho_{pp}(t) O^x_{pp}$ as an average over a probability distribution: Here, the matrix elements $\rho_{pp}(t)$ serve as the probabilities associated to the eigenvalues $O^x_{pp}$. Because the spectrum is dense, we can treat the distribution $\rho_{pp}$ as a continuous distribution $\rho(O^x)$ of the continuous variable $O^x$ \cite{PetrovYakovlev2012}. In our (finite-size) numerics, we obtain $\rho(O^x)$ as a histogram with appropriate bin sizes.


In Fig.~\ref{figOverhauserSpectrum}(a), we present the probability distribution of observable $O^x_{pp}$ after at $t=200\Tpulse$, $2000\Tpulse$, and $20000\Tpulse$, with couplings $a_j$ set such that the dephasing time $T^*$ has a realistic value of $T^*=3.16\ns$. The external magnetic field is set at $B_\mathrm{ext}=6\Tesla$. In the initial (thermal) state the distribution of Overhauser fields is approximately Gaussian. On the investigated time scale of $t=20000\Tpulse=264\mus$, the deviation from the initial distribution is hardly noticeable, and even smaller than the numerical noise caused by the discretization (binning). In order to extract the mode-locking effect, we examine the relative deviation $\rho_\mathrm{rel}(t)\equiv\rho_{pp}(t)/\rho_{pp}(0) - 1$, i.e., we divide the difference between the probability distribution at $t>0$ by the initial distribution by the latter. The result is shown in Figs.~\ref{figOverhauserSpectrum}(d). We find that the distance $\Delta \omega$ between the peaks approximately matches the pulsing rate, $\Delta\omega\approx 2\pi/\Tpulse=0.476\ns^{-1}\approx2\pi\times75.8\,\mathrm{MHz}$, so that we can attribute the observed effect to the synchronization to the pulses.

For a better illustration of the transformation to a peaked structure, we perform the same calculation with the coupling values enlarged by factors $\sqrt{10}$ and $10$, which shortens the dephasing time to $T^*= 1\ns$ and $T^*=0.316\ns$, respectively. The couplings are scaled uniformly, i.e., the ratios between the individual values are fixed. The results are exposed in Figs.~\ref{figOverhauserSpectrum}(b) and \ref{figOverhauserSpectrum}(c). The idea is that the deviation from the initial distribution grows much faster for these increased couplings. Comparison of the relative differences $\rho_\mathrm{rel}$ [Figs.~\ref{figOverhauserSpectrum}(e) and (f)] to Fig.~\ref{figOverhauserSpectrum}(d) shows that they are increased by factors of $10$ and $100$, respectively, compared to the realistic couplings. Thus, the growth rate of the peaks is roughly quadratic in the couplings.

Another difference between the distributions at $T^*=3.16\ns$, $1\ns$, and $0.316\ns$ is the number of peaks. The distance between the resonance peaks is unchanged, namely approximately equal to $2\pi/\Tpulse$, but the width of the distribution increases with decreasing dephasing time, so that more peaks are visible.

\subsection{Resonance condition}
\label{subsecResonance}
The question arises as to whether we can predict the location of the peaks in the Overhauser spectrum. We expect that whenever the system is mode locked, it admits a steady state, where the time evolution of the density matrix is periodic with a period of $\Tpulse$. In Appendix~\ref{appSteadyState} we demonstrate that, when we  consider the time evolution at zero order in the perturbation theory, we can find periodic solutions for arbitrary values of $\Omega\Tpulse$, so that we cannot single out a resonant value for the frequency $\Omega$. This property is due to the nature of the pulse, that maps any spin component perpendicular to the $z$ axis to a trion state, that subsequently decays in the Lindblad time evolution. Thus, the periodicity condition does not necessarily imply that an integer number of Larmor precessions fits inside the period $\Tpulse$.

As demonstrated by the peaks in the numerical results, the higher-order perturbative effects do not preserve this property. Due to the complicated structure of the perturbations (see Appendix~\ref{appFullPert}), we choose to avoid a direct derivation of the resonance condition through tedious algebra. Alternatively, we conjecture from the structure of the time evolution that the peaks correspond to an integer or to a half-integer number of Larmor oscillations, i.e., where exponentials of the form $\ee^{\ii\Omega\Tpulse}$ take the values $\pm 1$. The proposed condition is then tested empirically.

Two remarks are in place here. Firstly, we must take into account the second-order corrections $\Omega\pe2_{pp}$ in the frequencies. Whereas the value may be small, there is a large number of Larmor precessions in one period, so that the contribution $\Omega\pe2_{pp}\Tpulse$ adds up to a significant amount. Secondly, the trion decay leads to a small phase shift $\phi_\trion$ that is independent of the number of Larmor precessions between two pulses \cite{DuttEA2005,EconomouEA2005}. 
With those considerations, we conjecture our resonance condition to be
\begin{equation}\label{eqnResonanceCondition1}
 (\Omega\pe0_{pp} + \Omega\pe2_{pp})\Tpulse + \phi_\trion = n\pi,
\end{equation}
where $\Omega\pe0_{pp}=\lambda+\qf_p$ is the zeroth order frequency and $n$ is an integer, whose parity (even or odd) will be determined in due course. In the following, we investigate $\Omega\pe2_{pp}$ and $\phi_\trion$ in more detail.

The quadratic frequency shift is the second order perturbative correction to the eigenvalues,
\begin{equation}\label{eqnPertEigenvaluesTwoMain}
 \Omega_{pp}\pe2
  = \frac{1}{4}\sum_j a_j^2\left(\frac{\delta_{p^j,+}}{\lar+\qf_p-\frac{1}{2}a_j} + \frac{\delta_{p^j,-}}{\lar+\qf_p+\frac{1}{2}a_j}\right),
\end{equation}
where $\delta_{p^j,\pm} = 1$ if the $j$th spin of the basis vector $p$ is $\ket{\pm}$ and $0$ otherwise. (For details we refer to Appendix \ref{appFullPert}.) Due to the denominators  in \eqn\eqref{eqnPertEigenvaluesTwoMain} having an explicit dependence on $a_j$, there is no direct relation between the zeroth and second order frequency. However, if we approximate $\pm\tfrac{1}{2}a_j$ by its average value $\qf_p/N$, the denominators  can be approximated as $\lar+\qf_p(N-1)/N$,  eliminating the explicit dependence on $a_j$. In this approximation, the second order frequency shift is equal to
\begin{equation}\label{eqnPertEigenvaluesTwoApproxDiagonalMain}
 \Omega_{pp}\pe2=\frac{1}{4(\lar+\frac{N-1}{N}\qf_p)}\sum_j a_j^2,
\end{equation}
where the fixed value $\tfrac{1}{4}\sum_j a_j^2= \tfrac{1}{4}\AA$ is just a multiplicative prefactor. Substitution into \eqn\eqref{eqnResonanceCondition1} yields the resonance condition
\begin{equation}\label{eqnResonanceCondition2}
 \lambda + \qf_p +\frac{\AA/4}{\lar+\frac{N-1}{N}\qf_p}= \frac{n\pi - \phi_\trion}{\Tpulse}.
\end{equation}
This quadratic equation for $\qf_p=O^x_{pp}$ can be solved straightforwardly. For an intuitive understanding, we expand the solution in orders of $\AA$, which provides us with the peak positions
\begin{equation}\label{eqnResonanceConditionSolution}
 O^x(n)=\frac{n\pi - \phi_\trion}{\Tpulse} -\lar - \frac{N\AA/4}{\lar + (N-1)\frac{n\pi - \phi_\trion}{\Tpulse}} +\mathcal{O}(\mathcal{A}^2),
\end{equation}
for either even or odd integers $n$.

The physical reason behind the second-order frequency shift is the transverse component of the Overhauser field. The precession frequency of the central spin is proportional to the length of the total magnetic field $(B_\mathrm{ext}+B_\mathrm{O}^x,B_\mathrm{O}^y,B_\mathrm{O}^z)$, not just the longitudinal component \cite{BluhmEA2011}. The second order perturbation accounts for the transverse components of the Overhauser field. This geometrical argument also explains why the first non-trivial correction is of second order in the couplings.

The trion phase $\phi_\trion$ can be obtained from examination of the structure of the eigenvectors, e.g., as exhibited in the zeroth order time evolution \eqn\eqref{eqnEvoPertZero} and in Ref.~\cite{EconomouEA2005}. We analyze the Larmor precession through the expectation value $\avg{S^z}(t)$. Assuming a (post-pulse) initial state with $\avg{S^y}(0)=0$, we find
\begin{align}
 \avg{S^z}(t)&= \avg{S^z}(0)\cos\Omega t +\frac{\gamma^2 \rho_{\trion\trion}(0)}{\gamma^2+\Omega^2}\cos \Omega t \nonumber\\
 & \qquad{}+ \frac{\gamma\Omega\rho_{\trion\trion}(0)}{\gamma^2+\Omega^2}\sin \Omega t,
\end{align}
where we select one frequency $\Omega\equiv\Omega_{pp}$. With the familiar trigonometric identity $\cos(\Omega t+\phi)=\cos \Omega t\cos\phi-\sin\Omega t\sin\phi$, we obtain
\begin{align}
 R\sin\phi &= -\gamma\Omega\rho_{\trion\trion}(0)\\
 R\cos\phi &= (\Omega^2+\gamma^2)\avg{S^z}(0) + \gamma^2\rho_{\trion\trion}(0)\nonumber
\end{align}
for some positive constant $R$. Typically, the initial density matrix approaches $\avg{S^z}(0) = -\tfrac{1}{4}$, $\avg{S^y}(0) = 0$ and $\rho_{\trion\trion}=\tfrac{1}{2}$, which yields $\phi = \pi+ \arctan \gamma/\Omega$. The term $\pi$ comes from the fact that both $\sin\phi$ and $\cos\phi$ are negative. Subtracting the initial angle $\pi$ yields the trion phase 
\begin{equation}\label{eqnTrionPhase}
  \phi_\trion = \phi - \pi \approx \arctan \gamma/\Omega,
\end{equation}
where the right-most expression assumes the initial condition introduced above (cf.~ Refs.~\cite{DuttEA2005,EconomouEA2005}). In the limits considered here, $\Omega\approx \lar$ and $\gamma\ll\lar$, the trion phase is approximately equal to the ratio $\gamma/\lar$ between the trion decay rate and the Larmor frequency.

The physics behind the trion phase is the asymmetry between the central-spin up and down states while the trion decays \cite{DuttEA2005,EconomouEA2005}: The trion decays to spin up only, but is mixed into the down state as well by the Larmor precession. Because this mixing happens on a finite time scale, the trion amplitude has decreased in the time spin up is rotated to down. The asymmetry in mixing thus decreases if the Larmor precession is faster, consistent with the limit of small $\gamma/\Omega$ in  \eqn\eqref{eqnTrionPhase}. (The assumption $\gamma/\Omega\ll1$ is valid for all data presented in this work.) If the Larmor precession is slow compared to the trion decay, then the approximation in \eqn\eqref{eqnTrionPhase} is no longer valid, and other contributions appear that represent the effects of coherent trion recombination, known as spontaneously generated coherence \cite{DuttEA2005,EconomouEA2005}.

Finally, we empirically determine the parity of the integer $n$. We have explicitly calculated the solutions to the resonance condition \eqn\eqref{eqnResonanceConditionSolution} with the trion phase of \eqn\eqref{eqnTrionPhase}, and find that they line up well with the resonance peaks for odd $n$, as displayed in Figs.~\ref{figOverhauserSpectrum}(d)--(f) by the vertical lines. These contributions to the density matrix correspond to frequencies such that approximately a half-integer number of Larmor precessions fits into one period $\Tpulse$. 

Purely classical simulations with rather crude assumptions about the pulse and the trion decay also show the same dominant resonance behavior at half-integer precessions \cite{Hudepohl2016_masterthesis}, thereby supporting our findings here. Interestingly, there are indications \cite{PetrovYakovlev2012,Jaschke_privatecommunication} that the nuclear Zeeman effect, which is not included in our model, changes the parity from odd to even.

At present, we may only speculate why the half-integer number of Larmor precessions represent the more robust resonance condition. In Ref.~\cite{CarterEA2009}, a transition between the two parities has also been reported for off-resonant pulses, upon changing the sign of the detuning. There, the mechanism is understood through a nonzero $S^x$ polarization, which causes the transition rate for the nuclei from spin up to down to be different from that of the opposite process. For positive detuning, the system would diverge from the integer resonance condition into the half-integer one. Here, we have not considered detuned pulses, and we do not observe a significant nonzero spin expectation value along the magnetic axis. Whether the mechanism proposed in Ref.~\cite{CarterEA2009} also applies here is thus an interesting issue that is open for future research.

\subsection{Transverse components of the Overhauser field}

We have mentioned the effect of the transverse components of the Overhauser field on the resonance condition. In addition, with pulsing acting on the central spin in the $z$ direction, the question arises as to whether spin polarization is transferred to the nuclei. In that case, the transverse components $O^y$ and $O^z$ would attain nonzero expectation values after many pulses.

\begin{figure}
  \includegraphics{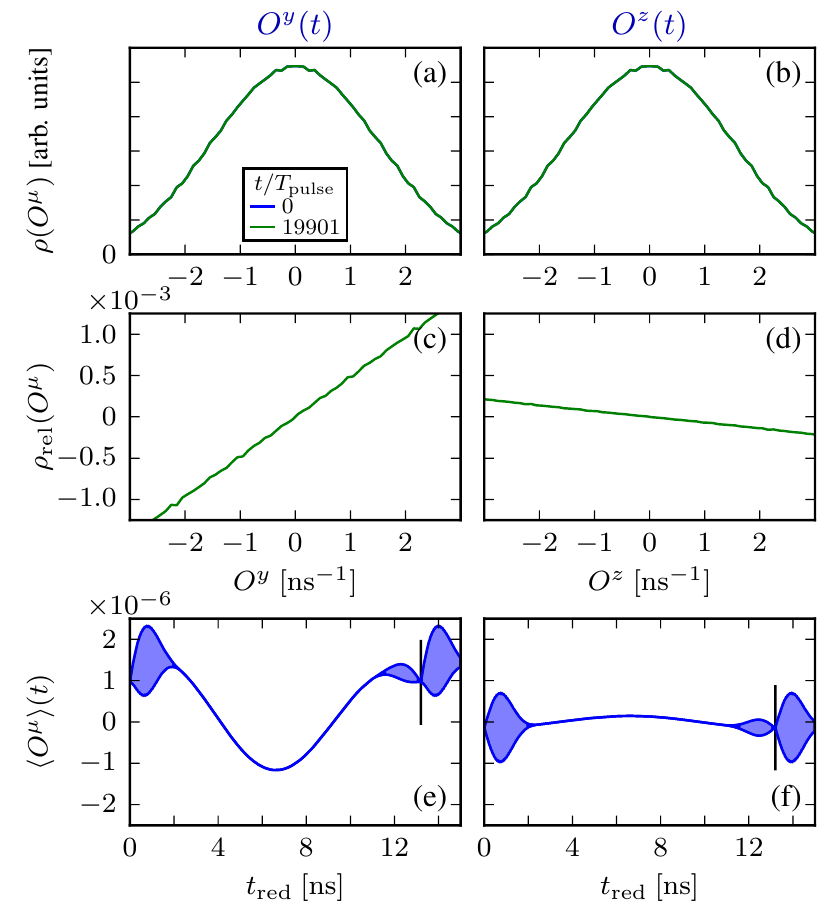}
  \caption{Behavior of the transverse components $O^\mu$ ($\mu=y,z$) after many pulses ($t=19901\Tpulse$). (a, b) Probability distributions of the observables in their respective eigenbasis. (c, d) Relative differences $\rho_\mathrm{rel}$ of the distribution at $t=19901\Tpulse$ with the initial one. (e, f) Time evolution $O^\mu(t)$ after $19901\Tpulse$, with $t_\mathrm{red} \equiv t- 19901\Tpulse$. The shaded regions indicate fast oscillations, with frequency close to the Larmor frequency $\lambda$. The nuclear couplings have been chosen such that $T^*=1\ns$.}
  \label{figTransverse}
\end{figure}

The expectation values $\avg{O^y}$ and $\avg{O^z}$ can be computed in the same way as the longitudinal counterpart. For extraction of probability distributions, analogously to $\avg{O^x}$ in \eqn\eqref{eqnExpectationValue}, the density matrix is first transformed into a  basis in which the observable is diagonal. For $O^y$ and $O^z$, the distributions are plotted in Fig.~\ref{figTransverse}(a) and (b). The distributions at large times are almost indistinguishable from the initial (Gaussian) distribution. The linear dependence of the relative difference $\rho_\mathrm{rel}$ [Figs.~\ref{figTransverse}(c) and (d)] indicates that the initial and final distributions are shifted slightly with respect to each other. These shifts are consistent with the finite values $\avg{O^y}(t)$ and $\avg{O^z}(t)$ at the moment of the pulse. [See Figs.~\ref{figTransverse}(e) and (f) for the time evolution of these expectation values. This time evolution is close to a steady state, i.e., approximately the same evolution repeats itself after every pulse.]

A striking difference to the distribution of $O^x$ is that the transverse components do not have the typical peak structure associated to mode locking. We furthermore observe that the width of the distribution remains almost invariant in all three directions. In other words, the dephasing time does not change over time.

The results also show the uncertainty in each of the three components $O^x$, $O^y$, and $O^z$. By virtue of the uncertainty principle, the three independent components of the Overhauser field cannot be determined with infinite precision, because they are defined from angular momentum operators which do not commute among each other. Hence, strictly speaking we cannot interpret a joint probability distribution of $O^x$, $O^y$, and $O^z$. However, the commutators scale as $\sum_j a_j^2$, so that the Overhauser field can be treated as almost classical in the limit of large $N$ \cite{StanekEA2014}. But here, we cannot apply a semiclassical approach, because the uncertainty defines a coarser frequency scale than the peak structure we desire to resolve.

\subsection{Mode-locking rate}
\label{subsecModeLockingRate}

Figure~\ref{figOverhauserSpectrum} shows that the formation of the peaks happens at a slow rate, which is expected to scale roughly as the square of the couplings $a_j$. This scaling law may also be understood from the following heuristic arguments. Firstly, the first-order perturbations to the entries of the density matrix can be understood as single-spin-flip processes with amplitudes in the order of $a_j/\lar$. Secondly, the distribution of Overhauser fields involves the diagonal entries $\rho_{pp}$. Due to the conservation of the trace of $\rho$, a change of $\rho_{pp}$ is linked to a change in $\rho_{p'p'}$. A transition between these matrix elements requires two spin flips, so that the corresponding amplitude is quadratic in $a_j/\lar$. We note the similarity to Fermi's golden rule, which is also second order in the perturbation.

\begin{figure}
  \includegraphics{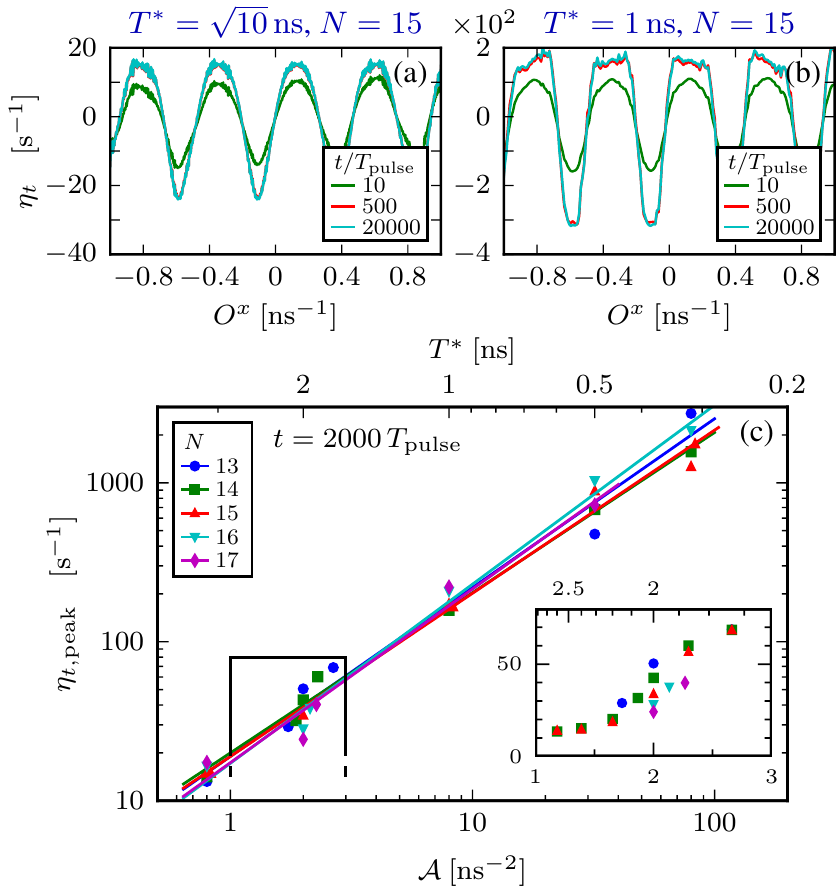}
  \caption{(a) Growth rate $\eta_t=\rho_\mathrm{rel}/t$ for couplings scaled such that $T^*=3.16\ns$ (a) or $T^*=1\ns$ (b). (c) Dependence of peak heights of $\rho_\mathrm{rel}$ on $\AA$ for different $N$. The lower horizontal axis shows $\AA$, and the upper horizontal axis the equivalent dephasing times $T^*$. For each $N$, we fit power laws $\propto\AA^\alpha$ with $\alpha\approx 1$, depicted as straight lines. In the inset, we zoom in on the regime around $T^*=2\ns$. The axes of the inset are linear.}
  \label{figOverhauserSpectrumScaling}
\end{figure}

In Figs.~\ref{figOverhauserSpectrumScaling}(a) and (b), we take the relative density distribution $\rho_\mathrm{rel}$ at time $t$, and divide it by $t$. The resulting quantity $\eta_t = \rho_{\mathrm{rel},t}/t$ is then compared for different times $t$. For large $t$, the curves for $\eta_t$ are almost identical, signifying linear growth of $\rho_\mathrm{rel}$ in time. In the short-time limit, in the order of $\sim 10$ pulses, $\eta_t$ tends to be smaller; the linear growth does not set in immediately. When the peaks become macroscopic ($\rho_\mathrm{rel}\gtrsim 0.1$), e.g., for $t=20000\,\Tpulse$ in Fig.~\ref{figOverhauserSpectrumScaling}(b), the growth accelerates, because it is exponential by nature. Within each pulse period, the entries of the density matrix increase or decrease by an amount proportional to the entries themselves. Of course, for small values of $\rho_\mathrm{rel}$, the exponential growth is indistinguishable from a linear dependence.

Comparing Figs.~\ref{figOverhauserSpectrumScaling}(a) and \ref{figOverhauserSpectrumScaling}(b), we find only small differences, except for the vertical scale being $10$ times larger in Fig.~\ref{figOverhauserSpectrumScaling}(b). This corroborates the earlier expectation that the peaks in the spectrum form at a rate proportional to the square of the couplings, or equivalently, inversely proportional to the square of the dephasing time $T^*$, given a fixed distribution of couplings up to an overall multiplicative factor.

In Fig.~\ref{figOverhauserSpectrumScaling}(c), we have compared the growth of the peaks in the spectrum for different numbers of nuclei $N$ and for different coupling strengths at a fixed point in time. The quantity of study is the peak value of $\eta_t$, averaged over the three peaks closest to $O^x=0$. In the main plot of Fig.~\ref{figOverhauserSpectrumScaling}(c), we have plotted on a double logarithmic scale in order to identify scaling laws of a power-law nature. Fitting power laws
$\eta_{t,\mathrm{peak}}\propto\AA^{\alpha}$
to the data for each individual $N$, we find exponents $\alpha=1.06\pm0.05$. Thus, the growth rate $\eta_{t,\mathrm{peak}}$ is approximately linear in $\AA$ or equivalently, $\propto1/(T^*)^2$. 

The scaling laws show a clear trend on a large range of coupling values, but closer inspection of the data points on a small range [see Fig.~\ref{figOverhauserSpectrumScaling}(c), inset] reveals a finer structure. Also, the scaling of the peak values as function of $N$ for a fixed coupling value is not definite: For $\sum_ja_j^2=2\ns^{-2}$, the peak growth decreases for increasing $N$, whereas it increases at, e.g., $\AA=0.8\ns^{-2}$ and $8\ns^{-2}$. We attribute this seemingly erratic behavior to discretization effects of the distribution of couplings. The coarse graining of this distribution leads to Overhauser spectra with different peak shapes, depending on the exact value of $\AA$. For large $N$, the distribution will be dense, and we expect the deviations from linear dependence to be smaller.

The linear fits over a broad range of values of $\AA$ eliminates this fine dependence. We compare the fit coefficients for different values of the number of nuclei $N$ in order to determine a scaling law in terms of this quantity. Based on the linear proportionality of the growth rate $\eta_t$ to $\AA$, where the latter is independent of $N$ \footnote{In this context, we normalize the distribution of couplings according to the value of $\AA$, which is set by the dephasing time $T^*$. Here, $N$ plays the role of a ``sampling resolution'' of the distribution. It should not be confused with its common interpretation as the effective number of nuclei in the system. In that interpretation, $\AA$ is a derived parameter that depends on the latter number.}, we would expect that $\eta_t$ is independent on $N$. However, other studies suggest the scaling law $\eta_t\propto N^{-1/2}$ \cite{Hudepohl2016_masterthesis}. Our data suggests a weak dependence of the growth rate on $N$, compatible with either of these cases, $\eta_t\propto N^0$ or $\eta_t\propto N^{-1/2}$. With the narrow range of system sizes studied here, and the relative large uncertainties in the fitting parameters, we are unable to determine which of both is more plausible.

We have also carried out a scaling analysis in terms of the external magnetic field strength $B_\mathrm{ext}$ as parameter.
Here, we find that the peak growth rate scales as $\eta \propto \lambda^{-2}$, where we recall that $\lambda=g\muB B_\mathrm{ext}/\hbar$. 
This scaling law follows from the structure of the perturbation theory, and confirms the idea mentioned before, namely, that the rate of mode locking scales as the square of the perturbation parameters $a_j/\lambda$. Experimental data confirms the qualitative behavior that a stronger magnetic field incurs faster dephasing, but quantitative measurements establishing the scaling law have not yet been performed \cite{GreilichBelykh_privatecommunication}.

\begin{figure}
  \includegraphics{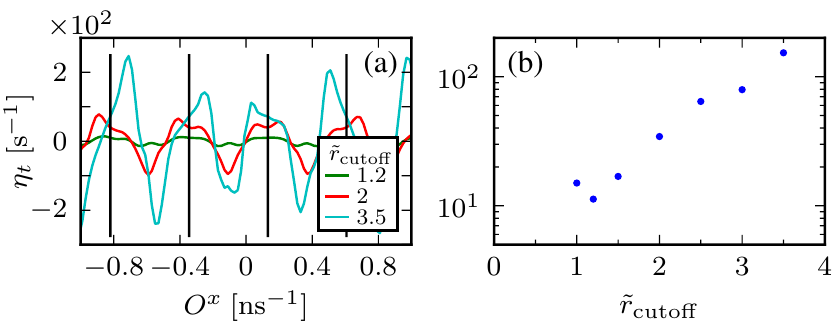}
  \caption{(a) Effect of the distribution of couplings on the peak structure. (b) Dependence of the peak growth rate on the normalized cutoff radius $\tilde{r}_\mathrm{cutoff}$. The squared sum of the couplings and the number of nuclei are fixed by $T^*=2\ns$ and $N=15$, respectively.}
  \label{figOverhauserSpectrumCutoff}
\end{figure}

The assumption that the distribution of couplings is fixed, is artificial in this numerical setting: for the small-$N$ numerics presented here, we have used a distribution of couplings based on a Gaussian wave function envelope, with a relatively small cutoff radius $\tilde{r}_\mathrm{cutoff}$, in order to prevent the largest coupling from dominating the nuclear dynamics. This construction cuts off the couplings with small values; thus, the \emph{physical} distribution of couplings would contain relatively more smaller couplings than the \emph{artificial} one. Figure~\ref{figOverhauserSpectrumCutoff}(a) shows that if we increase the cutoff value, the peak height of $\rho_\mathrm{rel}$ increases. In other words, by choosing the distribution of couplings with a small cutoff, we underestimate the growth rate $\eta$. In Fig.~\ref{figOverhauserSpectrumCutoff}(b), we plot the peak heights as a function of the cutoff values. The data suggests an increasing trend: The spectral peaks grow faster for a larger cutoff. Because the present data is strongly affected by the discretization due to the small value of $N$, we are not able to identify a specific dependence (e.g., exponential). Rigorous analysis of the dependence is left for future research.

On the other hand, we are also not capable of reaching the limit $\tilde{r}_\mathrm{cutoff}\to 0$ reliably. This limit corresponds to the box model, where all couplings have (almost) the same value. For small values of $N$, the distribution of all possible frequencies $\Omega_{pp}$ (as is the initial distribution of $O^x$) is no longer Gaussian, but peaked. In this situation we are unable to resolve the effect of mode locking. The box-model limit requires a different approach, namely, where the dynamics of the nuclear spins is treated collectively instead of each spin  individually \cite{BarnesEconomou2011}.

\subsection{Effect on the coherence}
\label{subsecCoherenceTimes}

\begin{figure}
  \includegraphics{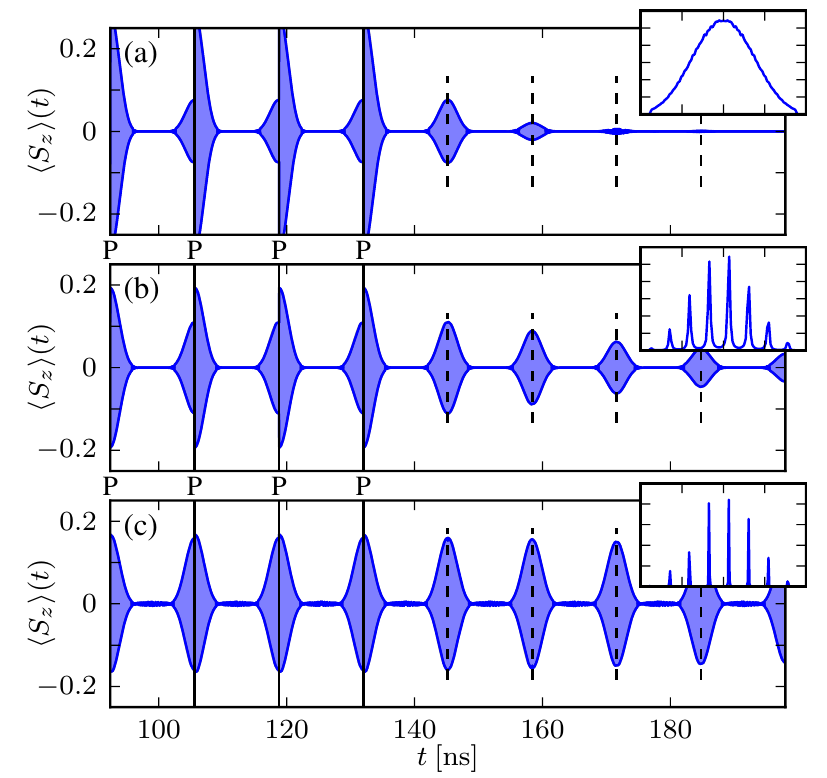}
  \caption{Amplitude of the central-spin Larmor oscillations after termination of periodic pulsing. Here, pulses (indicated by the vertical solid lines and the label P) are applied periodically every $13.2\ns$ until $t=132\ns$. The vertical dashed lines indicate $t=(132+13.2l)\ns$ ($l=1,2,\ldots$) where the revivals are located. The three panels differ in the amount of mode locking: (a) no mode locking, (b) weak mode locking (Lorentzian peaks with $w=0.05$), and (c) strong mode locking ($w=0.005$). The insets show the initial distribution of $O=\sum_j a_jI^x_j$. In all cases, $\AA=2\ns^{-2}$ ($T^*=2\ns$) and $N=15$. The shaded regions indicate fast oscillations.}
  \label{figRevivalComparison}
\end{figure}

As explained in the introduction, we distinguish two mechanisms that give rise to the revival effects in the central-spin Larmor oscillations. Firstly, as suggested by Fig.~\ref{figRevivalComparison}(a), we observe a revival effect that appears already after a few pulses, long before mode locking sets in. The mechanism for this revival is the nature of the pulse, combined with the incoherent decay of the trion. 
The system quickly converges to the steady state associated to this process. As demonstrated in Appendix~\ref{appSteadyState}, the steady-state expectation values of $\avg{S^z}(t)$ before and after the pulse are nonzero. In Fig.~\ref{figRevivalComparison}(a), we recover the pre- and post-pulse amplitudes of the steady state of approximately $0.077$ and $0.289$, respectively (see Appendix~\ref{appSteadyState}).

Unfortunately, observing a clear revival effect with mode-locked density matrices obtained after a long time evolution proves to be a challenge: For realistic values of the hyperfine couplings, the effect is too small, and for enlarged ones ($T^*\lesssim 0.5\ns$), the approximation errors add up, eventually leading to an unphysical density matrix with (small) negative diagonal entries. Instead, we artificially apply mode locking by multiplying the initial density matrix entries $\rho_{pq}$ by the function
\begin{equation}\label{eqnFocusingFunction}
  F(\Omega_{pq}) = \sum_{k=-\infty}^\infty l_w[(\Omega_{pq} - \omega_k)\Tpulse/2\pi],
\end{equation}
where $l_w(x) = w/[\pi(w^2+x^2)]$ designates a Lorentzian peak of width $w$, and $\omega_k= (2 k+1)\pi/\Tpulse + \phi_\trion$ are the resonant frequencies. These half-integer resonant frequencies coincide to high precision with those for the longitudinal Overhauser field $O^x$, as expressed by \eqns\eqref{eqnResonanceCondition2} and \eqref{eqnTrionPhase}. The Lorentzian peak shape should be interpreted as a generic example; other shapes will yield similar qualitative behavior \cite{*[{For a discussion about line shapes in NMR, see, e.g., }] [{}] WaeberEA2016}.

In the case of repeated pulsing, there appears to be no qualitative difference between the revivals in absence or in presence of mode locking. However, a remarkable difference arises in a pulse protocol where the pulsing is terminated at some moment, as demonstrated by Fig.~\ref{figRevivalComparison}. If we pulse until $t=l\Tpulse$, then there will be a clear revival at $t = (l+1)\Tpulse$. In absence of mode locking [see Fig.~\ref{figRevivalComparison}(a)], the subsequent revivals are significantly attenuated. In contrast, if the spectrum is mode locked [see Figs.~\ref{figRevivalComparison}(b) and (c)], the revivals at $t=(l+1)\Tpulse, (l+2)\Tpulse, \ldots$ are strong, and their amplitude decays slowly. The decay rate is determined by the amount of focusing: For narrower peaks, the revivals attenuate more slowly, and thus the coherence time is larger. The revival amplitudes decay exponentially as $\ee^{-t/\Tcoh}$, with a coherence time equal to $\Tcoh=\Tpulse/2\pi w$.
For the examples illustrated in Figs.~\ref{figRevivalComparison}(b) and (c), the coherence times are $42\ns$ and $420\ns$, respectively. The ratios of the amplitudes of subsequent peaks are $0.730$ and $0.969$, respectively.

In addition, the narrower the peaks, the more robust an ensemble will be against any statistical variation of the frequencies. In particular, the statistical variation in the frequencies $\Omega_{pq}$, caused by the slightly different $g$ factors of the individual quantum dots, does not alter the amplitudes of the revivals. Either without or with mode locking, the amplitude at the pulse times is unaffected, because both mechanisms filter the resonant contribution, which is independent of the Larmor frequency (or equivalently, of the $g$ factor). However, the additional statistical uncertainty in the distribution of frequencies $\Omega_{pq}$ leads to a shorter dephasing (and rephasing) time.

\subsection{Estimate of the minimal pulsing duration}

We combine the observations in Sec.~\ref{subsecModeLockingRate} in order to find an estimate of the time scale $\eta^{-1}$ at which the peaked structure sets in, by extrapolation to realistic parameter values. For concreteness, we assume an external magnetic field of $B_\mathrm{ext}=6\Tesla$ and a typical value of the dephasing time of $T^*\sim1\ns$ \cite{GreilichEA2006Science}. The effective number of nuclei is $N\sim10^5$.

From Fig.~\ref{figOverhauserSpectrumScaling}, we find a mode-locking rate of $\eta\sim10^2\,\mathrm{s}^{-1}$. As discussed in Sec.~\ref{subsecModeLockingRate}, the $N$-dependence is uncertain: both $\eta\propto N^0$ and $\eta\propto N^{-1/2}$ are plausible. In the latter case, $\eta$ is decreased by a factor of $\sim10^2$ for $N\sim10^5$. On the other hand, the low cutoff value $\tilde{r}_\mathrm{cutoff} = 2$ for the distribution of couplings leads to the mode-locking rate being underestimated. The data in Fig.~\ref{figOverhauserSpectrumCutoff} suggests that for realistic cutoff values, $\eta$ is increased slightly by up to roughly one order of magnitude.

The relation between the mode-locking rate $\eta$ and the necessary illumination time $\Tillum$ (duration during which the sample has to be pulsed) for a desired value of the coherence time $\Tcoh$, is obtained from the considerations in Sec.~\ref{subsecCoherenceTimes}. We equate the numerically obtained peak heights to those of $F(\Omega_{pq})$ given by \eqn\eqref{eqnFocusingFunction}, i.e.,
\begin{equation}\label{eqnPeakComparison}
  1+\rho_\mathrm{rel,peak}(\Tillum)
  = \coth \pi w
  \approx \frac{1}{\pi w}
  = \frac{2\Tcoh}{\Tpulse},
\end{equation}
where the approximation is valid if the peaks are sufficiently narrow ($w\ll 1$). If we assume that the peaks in the Overhauser spectrum grow exponentially as $1+\rho_\mathrm{rel,peak}(t)=\ee^{\eta t}$, we find $\Tillum = \eta^{-1}\ln (2\Tcoh/\Tpulse)$. For extremely long coherence times, e.g., as reported in Ref.~\cite{GreilichEA2007}, the ratio $\Tcoh/\Tpulse$ can be as large as $10^{10}$, for which $\Tillum\approx 24 \eta^{-1}$. This value should be considered as a lower bound: In a realistic scenario we expect that saturation will occur, i.e., that the exponential growth slows down when a high degree of mode locking is reached.

Combination of these observations leads to an estimate of the mode-locking rate of $\eta\sim10^2$--$10^3\,\mathrm{s}^{-1}$, assuming the scaling law $\eta\propto N^0$. Thus, the estimated minimal illumination time lies in the range of $0.1$--$1\,\mathrm{s}$. We re-emphasize that this value should be interpreted as a lower bound in view of the expected saturation effect discussed above.

\section{Discussion and conclusion}
\label{secConclusion}

Our estimate for the minimal pulsing duration that leads to the long coherence times reported in Ref.~\cite{GreilichEA2007} is in the order of $0.1$--$1\,\mathrm{s}$. In the experiments, the sample is illuminated for much longer, but it has not been investigated to what extent the long illumination time is required. As far as our knowledge reaches, the relation between the illumination time and the coherence time has not been investigated quantitatively.

Our analysis of the scaling in terms of the number of nuclei $N$ is uncertain, because we have access to a very limited number of values. Whereas for the estimation above we have assumed the mode-locking rate to be independent of $N$, our data is also compatible with the $N^{-1/2}$ scaling suggested by other studies \cite{Hudepohl2016_masterthesis}. With the latter scaling behavior, realistic values of $N$ imply a decrease of $\eta$ by $\sim10^2$, leading to an estimated minimal illumination time of $10$--$100\,\mathrm{s}$. 

Mode locking has also been addressed in studies that use (semi)classical approaches \cite{PetrovYakovlev2012,Jaschke_privatecommunication}. In these studies, a much faster growth of the peaks has been reported. We ascribe this difference to the loss of coherence at the pulses. For instance, in Ref.~\cite{PetrovYakovlev2012}, it is assumed that the pulses polarize the electron spin completely, regardless of the pre-pulse state. Thus, at each pulse the system is reset to a pure state, which leads to a much stronger effect of the resonance.

The qualitative distinction between the revival behavior in presence and absence of mode locking is recovered by experiments by Greilich \emph{et al.} \cite{GreilichEA2007,GreilichBelykh_privatecommunication}. In these measurements, the electron spin signal shows a revival effect on a fast time scale of ten pulses ($\sim 120\,\mathrm{ns}$). The pre-pulse amplitude is approximately $30 \%$ of the post-pulse amplitude, and the revivals die quickly after the pulses are switched off, which matches the behavior shown in Fig.~\ref{figRevivalComparison}(a). As of now, it is unknown whether the origin of this signal is the steady-state behavior as we describe here, or if it is a side effect of residual coherence between the measurements that are repeated every few microseconds. Secondly, the mode-locking effect [see Fig.~\ref{figRevivalComparison}(c)] requires a pulsing duration in the order of seconds or beyond, and the coherence effect is retained on even longer time scales. Experimental results have also confirmed that the pre- and post-pulse amplitude have (almost) the same value in this case \cite{GreilichEA2007}.

It should be emphasized that we have chosen the philosophy of analyzing a minimal model that clarifies the phenomenon of mode locking. Hereby, we have neglected several interactions known to have a quantitative effect on the results. In particular, it has been suggested that the nuclear Zeeman effect, absent in our model, leads to a significant decrease in the mode-locking rate \cite{PetrovYakovlev2012,Jaschke_privatecommunication}. Further interactions that affect the nuclear dynamics are the quadrupolar interaction of the nuclei (in case they are  considered as spin-$\tfrac{3}{2}$ particles) \cite{SinitsynEA2012,HackmannEA2015,BechtoldEA2015,SokolovEA2016}, the dipole-dipole interaction between nuclei \cite{AuerEA2009}, and anisotropy of the dipolar hyperfine interaction or of the $g$ factors (in case of a hole central spin rather than an electron) \cite{FischerEA2008,TestelinEA2009,ChekhovichEA2013,HackmannAnders2014,HackmannEA2015,VidalEA2016,BelykhEA2016}. The present framework of perturbation theory could be extended with these additional interactions with relatively small effort. The present  framework also enables us to investigate the effect of the pulse action, in particular, how off-resonant pulses give rise to nuclear spin polarization in the magnetic-field direction \cite{CarterEA2009,BarnesEconomou2011}. An extensive analysis of additional interactions and of other pulse types lies beyond the scope of this work.

The perturbative method also has its limitations. For realistic couplings, the effect of mode locking becomes visible only for unfeasibly long times. On the other hand, if the couplings are artificially increased, the errors (being quadratic in the couplings) grow much more rapidly, so that the resulting density matrices become unphysical before we reach times for which the focusing effect becomes significant. For more precise estimates and a longer time interval for the evolution, further development of our methods may be required. For example, we could eliminate the error from not including multi-spin-flip processes within a single pulsing period, which arises due to the perturbation theory being of first order in the eigenvectors. Dividing the pulsing interval into multiple sub-intervals alleviates this problem to some extent, but may also introduce additional truncation errors which may become significant if the sub-intervals are too short.

Alternative promising approaches towards calculation of the central-spin-model dynamics have been proposed, such as diagrammatic perturbation theory \cite{CoishEA2010}, exact time evolution \cite{BarnesEA2012}, density matrix renormalization group (DMRG) methods \cite{StanekEA2013,StanekEA2014,GravertEA2016}, Monte Carlo methods \cite{FaribaultSchuricht2013PRL,*FaribaultSchuricht2013PRB}, and approaches employing conserved quantities \cite{ChenEA2007,UhrigEA2014}. Each of these methods should be scrutinized as to how well they are suited and capable of capturing the mode locking effect. One essential requirement is that sufficient information on the nuclear configuration is carried over from one pulse to the next. Methods which treat the Overhauser field naively as a classical variable (e.g., the expectation value only) and violate this requirement, are by nature unable to capture the physics of mode locking correctly. 

\acknowledgments

We acknowledge financial support from the Deutsche Forschungsgemeinschaft  and the Russian Foundation of Basic Research in the framework of ICRC TRR 160. We are grateful to A. Greilich, V. V. Belykh, N. J\"aschke, and M. M. Glazov for inspiring discussions.

\appendix

\section{Full perturbative results}
\label{appFullPert}
In this Appendix, we provide an overview of the perturbative results for clarification and for reference. We first review the known solution of finding the eigenvalues and eigenvectors of the zeroth order Liouville operator $\Li\pe0$ \cite{EconomouBarnes2014,MerkulovEA2002}. Subsequently, we build the higher order perturbations based on top of this result.

For studying the perturbative expansion of the Liouville operator $\Li$, we represent in a matrix language where it is encoded as a $D^2\times D^2$ matrix, where $D=\dim \mathcal{H}=3\times2^N$ is the Hilbert-space dimension. Due to the decay term [see \eqn\eqref{eqnLi0}], it cannot be represented as a $D\times D$ matrix.

The zeroth order $\Li\pe0$ [\eqn\eqref{eqnLi0}] has been chosen to be block diagonal in the nuclear degrees of freedom. Working in the $\hat{x}$ basis, we can write the $(p,q)$ block in the central-spin basis $\{\ket{+}\bra{+},\ket{-}\bra{-},\ket{+}\bra{-},\ket{-}\bra{+},\ket{\trion}\bra{\trion},$ $\ket{+}\bra{\trion},\ket{\trion}\bra{+},$ $\ket{-}\bra{\trion},\ket{\trion}\bra{-}\}$ as the $9\times9$ matrix
\begin{widetext}
\begin{equation}\label{eqnLi0pq}
 \Li\pe0_{pq} = \begin{pmatrix}
             -\ii\omega_{pq} & & & & \half\gamma\\
             & \ii \omega_{pq} & & & \half\gamma\\
             & &-\ii \Omega_{pq} & & \half\gamma\\
             & & & \ii \Omega_{pq} & \half\gamma\\
             & & & & -\gamma\\
             & & & & &-\ii\epsilon^+_p-\half\gamma\\
             & & & & & & \ii\epsilon^+_q-\half\gamma\\
             & & & & & & &-\ii\epsilon^-_p-\half\gamma\\
             & & & & & & & & \ii\epsilon^-_q-\half\gamma\\
            \end{pmatrix},
\end{equation}
\end{widetext}
where the zero entries have been left blank. The entries on the diagonal are given in terms of $\omega_{pq}$ and $\Omega_{pq}$ as given by \eqn\eqref{eqnOmegas}, and of
\begin{equation}\label{eqnEpsilonPlusMinus}
  \epsilon^\pm_p =\pm\half(\lar + \qf^p)-\ET/\hbar,
\end{equation}
where $\qf^p$ is defined by \eqn\eqref{eqnQF}.

The matrix $\Li\pe0_{pq}$ itself has an internal block structure: There is one $5\times5$ block consisting of the degrees of freedom $\ket{+}\bra{+}$, $\ket{-}\bra{-}$, $\ket{+}\bra{-}$, $\ket{-}\bra{+}$, and $\ket{\trion}\bra{\trion}$, which we will refer to as the spin-spin/trion-trion (SS/TT) sector. The spin-trion/trion-spin (ST/TS) components $\ket{+}\bra{\trion}$, $\ket{\trion}\bra{+}$, $\ket{-}\bra{\trion}$, and $\ket{\trion}\bra{-}$ are all uncoupled.

We continue with the diagonalization of the zeroth order. In the spin-spin sector, we simply have the eigenvectors $\ket{+}\bra{+}\otimes\ket{p}\bra{q}$, $\ket{-}\bra{-}\otimes\ket{p}\bra{q}$, $\ket{+}\bra{-}\otimes\ket{p}\bra{q}$, $\ket{-}\bra{+}\otimes\ket{p}\bra{q}$. The trion-trion eigenvector is $r^{\trion\trion}\otimes\ket{p}\bra{q}$, with
\begin{align}
  r^{\trion\trion} &= \ket{\trion}\bra{\trion}
                    + \frac{\Half\ii\gamma}{-\omega_{pq}-\ii\gamma}\ket{+}\bra{+} 
                    + \frac{\Half\ii\gamma}{\omega_{pq}-\ii\gamma}\ket{-}\bra{-}\nonumber\\ 
                   &\qquad{}+ \frac{\Half\ii\gamma}{-\Omega_{pq}-\ii\gamma}\ket{+}\bra{-}
                    + \frac{\Half\ii\gamma}{\Omega_{pq}-\ii\gamma}\ket{-}\bra{+}.
\label{eqnRightEvecTT}
\end{align}
This eigenvector couples the trion-trion and the spin-spin degrees of freedom together to form the SS/TT sector. For the ST/TS sector, the eigenvectors are $\ket{+}\bra{\trion}\otimes\ket{p}\bra{q}$, $\ket{\trion}\bra{+}\otimes\ket{p}\bra{q}$, $\ket{-}\bra{\trion}\otimes\ket{p}\bra{q}$, and $\ket{\trion}\bra{-}\otimes\ket{p}\bra{q}$. The respective eigenvalues are $-\ii\omega_{pq}$, $\ii\omega_{pq}$, $-\ii\Omega_{pq}$, $\ii\Omega_{pq}$ and $-\gamma$ for the SS/TT sector, where $\omega_{pq}$ and $\Omega_{pq}$ are given by \eqn\eqref{eqnOmegas}, and represent the oscillating modes. The purely real eigenvalue $-\gamma$ encodes the trion decay. The eigenvalues associated to the ST/TS sector are $-\half\gamma-\ii\epsilon^+_p$, $-\half\gamma+\ii\epsilon^+_q$, $-\half\gamma-\ii\epsilon^-_p$, and $-\half\gamma+\ii\epsilon^-_q$,  
which are mixed real and imaginary, and thus represent decaying oscillations.

In matrix language, the Liouville operator is diagonalized as $\Li = RDL$, where $D$ is the diagonal matrix of eigenvalues, $R$ the matrix of right eigenvectors, and $L=R^{-1}$ the matrix of left eigenvectors. The matrix $\Li$ being non-hermitian means that the eigenvalues are generally complex, and $L=R^{-1}\not=R^\dagger$. Similarly, we have $\Li\pe0 = R\pe0 D\pe0 L\pe0$, with 
\begin{align}
 D\pe0_{pq} &= \diag(-\ii\omega_{pq}, \ii\omega_{pq}, -\ii\Omega_{pq}, \ii\Omega_{pq}, -\gamma,\\
  &\qquad -\half\gamma-\ii\epsilon^+_p, -\half\gamma+\ii\epsilon^+_q, -\half\gamma-\ii\epsilon^-_p,-\half\gamma+\ii\epsilon^-_q)\nonumber
\end{align}
as the diagonal matrix of eigenvalues and with
\begin{equation}
 R\pe0_{pq;p'q'}=\delta_{pp'}\delta_{qq'}
 \begin{pmatrix}
   1& & & & \frac{\ii\gamma/2}{-\omega_{pq}-\ii\gamma}\\
    &1& & & \frac{\ii\gamma/2}{\omega_{pq}-\ii\gamma}\\
    & &1& & \frac{\ii\gamma/2}{-\Omega_{pq}-\ii\gamma}\\
    & & &1& \frac{\ii\gamma/2}{\Omega_{pq}-\ii\gamma}\\
    & & & & 1\\
    & & & &  & 1\\
    & & & &  & & 1\\
    & & & &  & & & 1\\
    & & & &  & & & & 1
 \end{pmatrix}
\end{equation}
and
\begin{equation}
 L\pe0_{pq;p'q'}=\delta_{pp'}\delta_{qq'}
 \begin{pmatrix}
   1& & & & \frac{-\ii\gamma/2}{-\omega_{pq}-\ii\gamma}\\
    &1& & & \frac{-\ii\gamma/2}{\omega_{pq}-\ii\gamma}\\
    & &1& & \frac{-\ii\gamma/2}{-\Omega_{pq}-\ii\gamma}\\
    & & &1& \frac{-\ii\gamma/2}{\Omega_{pq}-\ii\gamma}\\
    & & & & 1\\
    & & & &  & 1\\
    & & & &  & & 1\\
    & & & &  & & & 1\\
    & & & &  & & & & 1
 \end{pmatrix},
\end{equation}
as the matrices of right and left eigenvectors, respectively. The latter two are related by inversion, $L\pe0 = (R\pe0)^{-1}$. The right and left eigenvectors are represented by the columns of $R$ and the rows of $L$, respectively. The Kronecker deltas indicate that the matrices are diagonal in the nuclear indices, i.e., these $9\times9$ matrices are the blocks for a single value of the nuclear indices ($p,q$).

The time evolution at zero order $\ee^{t\Li\pe0} = R\pe0\ee^{tD\pe0}L\pe0$ is then calculated straightforwardly as
\begin{widetext}
\begin{align}
 \ee^{t\Li\pe0} 
  =\sum_{pq}\biggl[{}&{} \ee^{ \ii t \omega_{pq} }\ket{pq;--}\bra{pq;--}
  +  \ee^{-\ii t \omega_{pq} }\ket{pq;++}\bra{pq;++}
  + \ee^{ \ii t \Omega_{pq} }\ket{pq;-+}\bra{pq;-+}
  +  \ee^{-\ii t \Omega_{pq} }\ket{pq;+-}\bra{pq;+-} \nonumber \\
  &+ \ee^{-\Half\gamma t - \ii \epsilon^+_p t} \ket{pq;+\trion}\bra{pq;+\trion} 
  +  \ee^{-\Half\gamma t + \ii \epsilon^+_q t} \ket{pq;\trion+}\bra{pq;\trion+}\nonumber\\
  &+  \ee^{-\Half\gamma t - \ii \epsilon^-_p t} \ket{pq;-\trion}\bra{pq;-\trion} 
  + \ee^{-\Half\gamma t + \ii \epsilon^-_q t} \ket{pq;\trion-}\bra{pq;\trion-}
  +\ee^{- \gamma  t}\ket{pq;\trion\trion}\bra{pq;\trion\trion}\nonumber\\
  &+\frac{\Half\ii\gamma}{-\omega_{pq}-\ii \gamma } \left(\ee^{- \gamma  t}- \ee^{-\ii t \omega_{pq} }\right)\ket{pq;++}\bra{pq;\trion\trion} 
  +\frac{\Half\ii\gamma}{ \omega_{pq}-\ii \gamma } \left(\ee^{- \gamma  t}- \ee^{ \ii t \omega_{pq} }\right)\ket{pq;--}\bra{pq;\trion\trion} \nonumber \\
  &+\frac{\Half\ii\gamma}{-\Omega_{pq}-\ii \gamma } \left(\ee^{- \gamma  t}- \ee^{-\ii t \Omega_{pq}}\right)\ket{pq;+-}\bra{pq;\trion\trion} 
  +\frac{\Half\ii\gamma}{ \Omega_{pq}-\ii \gamma } \left(\ee^{- \gamma  t}- \ee^{ \ii t \Omega_{pq}}\right)\ket{pq;-+}\bra{pq;\trion\trion}\biggr].
  \label{eqnEvoPertZero}
\end{align}
\end{widetext}

The first order correction to the time evolution follows from expansion of the matrices $D=D\pe0+D\pe1+\ldots$, $R=R\pe0+R\pe1+\ldots$, and $L=L\pe0+L\pe1+\ldots$ into orders of $\Li\pe1$, and subsequent substitution into $\Li = RDL$, with the condition that $\Li\pe0 = R\pe0 D\pe0 L\pe0$. The first order perturbations of the eigenvalues $\mu_\alpha$ of $\Li$ are equal to $
\mu\pe1_\alpha = \bramidket{l_\alpha\pe0}{\Li\pe1}{r_\alpha\pe0}$, where $\bra{l_\alpha\pe0}$ and $\ket{r\pe0_\alpha}$ are the left and right eigenvectors, respectively, of the zeroth order problem, associated to eigenvalue $\mu\pe0$. The spin flip in $H\pe1$ [\eqn\eqref{eqnH1}] maps each eigenvector either to zero or to a perpendicular eigenspace, so that $\mu\pe1_\alpha$ vanishes. Thus, the first order perturbation of the eigenvalues is trivial, i.e., $D\pe1=0$.

The first order perturbations to the right eigenvectors follow from
\begin{equation}\label{eqnPertEigenvectorsOne}
  \ket{r\pe1_\alpha} = \sum_{\beta:\mu\pe0_\beta\not=\mu\pe0_\alpha}
                       \ket{r\pe0_\beta}\frac{\bramidket{l\pe0_\beta}{\Li\pe1}{r\pe0_\alpha}}{\mu\pe0_\alpha-\mu\pe0_\beta}.
\end{equation}
As the operator $\Li\pe1$ involves exactly one nuclear spin flip, the perturbation to an eigenvector $\ket{r\pe0}$ with nuclear indices $(p,q)$ has contributions living in the nuclear spaces $(p',q)$ and $(p,q')$, where the first or second index, respectively, is raised or lowered for one nucleus. (In total, this constitutes $2N$ possibilities.)

The first-order eigenvector corrections, as encoded by $R\pe1$ and $L\pe1=-L\pe0 R\pe1 L\pe0$, are quite lengthy, hence we only provide the resulting correction to the time evolution, and leave out the intermediate steps. The expansion of the time evolution up to first order is given by
\begin{align}
  \ee^{t\Li} &= R\ee^{tD}L \approx R\pe0\ee^{tD\pe0}L\pe0
\\&\hspace{20mm}+R\pe0\ee^{tD\pe0}L\pe1+R\pe1\ee^{tD\pe0}L\pe0+\ldots,\nonumber
\end{align}
where the first term on the right-hand side is the zeroth order [\eqn\eqref{eqnEvoPertZero}] and the two following terms constitute the first order. The latter can be written as sums over the nuclear configurations $(p,q)$ and over the nuclei $j$,
\begin{align}
  R\pe0\ee^{tD\pe0}L\pe1  &= \frac{1}{2}\sum_{pq}\sum_{j=1}^N a_j \left(\mathcal{E}^{01}_{j;pq}+\mathcal{F}^{01}_{j;pq}+\mathcal{G}^{01}_{j;pq}\right),\nonumber\\
  R\pe1\ee^{tD\pe0}L\pe0  &= \frac{1}{2}\sum_{pq}\sum_{j=1}^N a_j \left(\mathcal{E}^{10}_{j;pq}+\mathcal{F}^{10}_{j;pq}+\mathcal{G}^{10}_{j;pq}\right),
\label{eqnEvoPertOne}
\end{align}
where
\begin{widetext}
\begin{align}
 \mathcal{E}^{01}_{j;pq}=
   -& \frac{\ee^{\ii t (-\omega +\frac{1}{2}a_j)}}{\frac{1}{2}a_j-\omega -\Omega }\ket{\pminusq;++}\bra{\pplusq;-+} 
  -\frac{\frac{1}{2}\ii \gamma  \ee^{\ii t (-\omega +\frac{1}{2}a_j)}}{(-\frac{1}{2}a_j+ \ii \gamma +\omega ) (\frac{1}{2}a_j-\omega -\Omega )}\ket{\pminusq;++}\bra{\pplusq;\trion\trion} \nonumber\\ 
  -& \frac{\ee^{\ii t (\omega +\frac{1}{2}a_j)}}{\frac{1}{2}a_j+\omega +\Omega }\ket{\pplusq;--}\bra{\pminusq;+-} 
  -\frac{\frac{1}{2}\ii \gamma  \ee^{\ii t (\omega +\frac{1}{2}a_j)}}{(-\frac{1}{2}a_j+ \ii \gamma -\omega ) (\frac{1}{2}a_j+\omega +\Omega )}\ket{\pplusq;--}\bra{\pminusq;\trion\trion} \nonumber\\ 
  -& \frac{\ee^{\ii t (-\omega -\frac{1}{2}a_j)}}{\frac{1}{2}a_j+\omega -\Omega }\ket{\pqminus ;++}\bra{\pqplus;+-} 
  -\frac{\frac{1}{2}\ii \gamma  \ee^{\ii t (-\omega -\frac{1}{2}a_j)}}{(\frac{1}{2}a_j+ \ii \gamma +\omega ) (\frac{1}{2}a_j+\omega -\Omega )}\ket{\pqminus ;++}\bra{\pqplus;\trion\trion} \nonumber\\ 
  -& \frac{\ee^{\ii t (\omega -\frac{1}{2}a_j)}}{\frac{1}{2}a_j-\omega +\Omega }\ket{\pqplus ;--}\bra{\pqminus;-+} 
  -\frac{\frac{1}{2}\ii\gamma \ee^{\ii t (\omega -\frac{1}{2}a_j)}}{(\frac{1}{2}a_j+ \ii \gamma - \omega ) (\frac{1}{2}a_j-\omega +\Omega )}\ket{\pqplus ;--}\bra{\pqminus;\trion\trion}
\label{eqnEvoPertOne01ssttSmallOmega},
\end{align}

\begin{align}
 \mathcal{E}^{10}_{j;pq}=
  &\frac{\ee^{\ii t \omega }}{\frac{1}{2}a_j-\omega -\Omega }\ket{\pminusq;+-}\bra{\pplusq;--} 
  -\frac{\Half\ii \gamma }{-\omega + \ii \gamma } \left(\frac{\ee^{- \gamma  t}}{\frac{1}{2}a_j- \ii \gamma -\Omega }-\frac{\ee^{\ii t \omega }}{\frac{1}{2}a_j-\omega -\Omega }\right)\ket{\pminusq;+-}\bra{\pplusq;\trion\trion} \nonumber\\
  +&\frac{\ee^{-\ii t \omega }}{\frac{1}{2}a_j+\omega +\Omega }\ket{\pplusq;-+}\bra{\pminusq;++} 
  -\frac{\Half\ii \gamma }{\omega + \ii \gamma } \left(\frac{\ee^{- \gamma  t}}{\frac{1}{2}a_j- \ii \gamma +\Omega }-\frac{\ee^{-\ii t \omega }}{\frac{1}{2}a_j+\omega +\Omega }\right)\ket{\pplusq;-+}\bra{\pminusq;\trion\trion} \nonumber\\
  +&\frac{\ee^{\ii t \omega }}{\frac{1}{2}a_j+\omega -\Omega }\ket{\pqminus ;-+}\bra{\pqplus;--} 
  -\frac{\Half\ii \gamma }{-\omega + \ii \gamma } \left(\frac{\ee^{- \gamma  t}}{\frac{1}{2}a_j+ \ii \gamma -\Omega }-\frac{\ee^{\ii t \omega }}{\frac{1}{2}a_j+\omega -\Omega }\right)\ket{\pqminus ;-+}\bra{\pqplus;\trion\trion} \nonumber\\
  +&\frac{\ee^{-\ii t \omega }}{\frac{1}{2}a_j-\omega +\Omega }\ket{\pqplus ;+-}\bra{\pqminus;++} 
  -\frac{\Half\ii \gamma }{\omega + \ii \gamma } \left(\frac{\ee^{- \gamma  t}}{\frac{1}{2}a_j+ \ii \gamma +\Omega }-\frac{\ee^{-\ii t \omega }}{\frac{1}{2}a_j-\omega +\Omega }\right)\ket{\pqplus ;+-}\bra{\pqminus;\trion\trion} 
\label{eqnEvoPertOne10ssttSmallOmega},
\end{align}
\begin{align}
 \mathcal{F}^{01}_{j;pq}=
  -& \frac{\ee^{\ii t (-\Omega +\frac{1}{2}a_j)}}{\frac{1}{2}a_j-\omega -\Omega }\ket{\pminusq;+-}\bra{\pplusq;--} 
  -\frac{\frac{1}{2}\ii \gamma  \ee^{\ii t (-\Omega +\frac{1}{2}a_j)}}{(-\frac{1}{2}a_j+ \ii \gamma +\Omega ) (\frac{1}{2}a_j-\omega -\Omega )}\ket{\pminusq;+-}\bra{\pplusq;\trion\trion} \nonumber\\ 
  -& \frac{\ee^{\ii t (\Omega +\frac{1}{2}a_j)}}{\frac{1}{2}a_j+\omega +\Omega }\ket{\pplusq;-+}\bra{\pminusq;++} 
  -\frac{\frac{1}{2}\ii \gamma  \ee^{\ii t (\Omega +\frac{1}{2}a_j)}}{(-\frac{1}{2}a_j+ \ii \gamma -\Omega ) (\frac{1}{2}a_j+\omega +\Omega )}\ket{\pplusq;-+}\bra{\pminusq;\trion\trion} \nonumber\\ 
  -& \frac{\ee^{\ii t (\Omega -\frac{1}{2}a_j)}}{\frac{1}{2}a_j+\omega -\Omega }\ket{\pqminus ;-+}\bra{\pqplus;--} 
  -\frac{\frac{1}{2}\ii\gamma  \ee^{\ii t (\Omega -\frac{1}{2}a_j)}}{(\frac{1}{2}a_j+ \ii \gamma - \Omega ) (\frac{1}{2}a_j+\omega -\Omega )}\ket{\pqminus ;-+}\bra{\pqplus;\trion\trion}  \nonumber\\ 
  -& \frac{\ee^{\ii t (-\Omega -\frac{1}{2}a_j)}}{\frac{1}{2}a_j-\omega +\Omega }\ket{\pqplus ;+-}\bra{\pqminus;++}
   -\frac{\frac{1}{2}\ii \gamma  \ee^{\ii t (-\Omega -\frac{1}{2}a_j)}}{(\frac{1}{2}a_j+ \ii \gamma +\Omega ) (\frac{1}{2}a_j-\omega +\Omega )}\ket{\pqplus ;+-}\bra{\pqminus;\trion\trion}
\label{eqnEvoPertOne01ssttLargeOmega},
\end{align}
and
\begin{align}
\mathcal{F}^{10}_{j;pq}=
   & \frac{\ee^{\ii t \Omega }}{\frac{1}{2}a_j-\omega -\Omega }\ket{\pminusq;++}\bra{\pplusq;-+} 
  -\frac{\Half\ii \gamma }{-\Omega + \ii \gamma } \left(\frac{\ee^{- \gamma  t}}{\frac{1}{2}a_j- \ii \gamma -\omega }-\frac{\ee^{\ii t \Omega }}{\frac{1}{2}a_j-\omega -\Omega }\right)\ket{\pminusq;++}\bra{\pplusq;\trion\trion} \nonumber\\
  +& \frac{\ee^{-\ii t \Omega }}{\frac{1}{2}a_j+\omega +\Omega }\ket{\pplusq;--}\bra{\pminusq;+-} 
  -\frac{\Half\ii \gamma }{\Omega + \ii \gamma } \left(\frac{\ee^{- \gamma  t}}{\frac{1}{2}a_j- \ii \gamma +\omega }-\frac{\ee^{-\ii t \Omega }}{\frac{1}{2}a_j+\omega +\Omega }\right)\ket{\pplusq;--}\bra{\pminusq;\trion\trion} \nonumber\\
  +& \frac{\ee^{-\ii t \Omega }}{\frac{1}{2}a_j+\omega -\Omega }\ket{\pqminus ;++}\bra{\pqplus;+-} 
  -\frac{\Half\ii \gamma }{\Omega + \ii \gamma } \left(\frac{\ee^{- \gamma  t}}{\frac{1}{2}a_j+ \ii \gamma +\omega }-\frac{\ee^{-\ii t \Omega }}{\frac{1}{2}a_j+\omega -\Omega }\right)\ket{\pqminus ;++}\bra{\pqplus;\trion\trion} \nonumber\\
  +& \frac{\ee^{\ii t \Omega }}{\frac{1}{2}a_j-\omega +\Omega }\ket{\pqplus ;--}\bra{\pqminus;-+ }
  -\frac{\Half\ii \gamma }{-\Omega + \ii \gamma } \left(\frac{\ee^{- \gamma  t}}{\frac{1}{2}a_j+ \ii \gamma -\omega }-\frac{\ee^{\ii t \Omega }}{\frac{1}{2}a_j-\omega +\Omega }\right)\ket{\pqplus ;--}\bra{\pqminus;\trion\trion},
\label{eqnEvoPertOne10ssttLargeOmega}
\end{align}
contain all couplings within the SS/TT sector with frequencies close to $\omega_{pq}$ and $\Omega_{pq}$, respectively, and
\begin{align}
 \mathcal{G}^{01}_{j;pq}=
  -&\frac{\ee^{-\Half\gamma t - \ii (\epsilon_p^+-\frac{1}{2}a_j) t }}{\frac{1}{2}a_j-\omega -\Omega }\ket{\pminusq;+\trion}\bra{\pplusq;-\trion} 
  -\frac{\ee^{-\Half\gamma t - \ii (\epsilon_p^--\frac{1}{2}a_j) t }}{\frac{1}{2}a_j+\omega +\Omega }\ket{\pplusq;-\trion}\bra{\pminusq;+\trion}\nonumber\\ 
  -&\frac{\ee^{-\Half\gamma t + \ii (\epsilon^q_--\frac{1}{2}a_j) t }}{\frac{1}{2}a_j+\omega -\Omega }\ket{\pqminus ;\trion+}\bra{\pqplus;\trion-}
  -\frac{\ee^{-\Half\gamma t + \ii (\epsilon^q_--\frac{1}{2}a_j) t }}{\frac{1}{2}a_j-\omega +\Omega }\ket{\pqplus ;\trion-}\bra{\pqminus;\trion+}
\label{eqnEvoPertOne01stst}
\end{align}
and
\begin{align}
 \mathcal{G}^{10}_{j;pq}=
   &\frac{\ee^{-\Half\gamma t - \ii \epsilon^-_p t}}{\frac{1}{2}a_j-\omega -\Omega }\ket{\pminusq;+\trion}\bra{\pplusq;-\trion} 
  + \frac{\ee^{-\Half\gamma t - \ii \epsilon^+_p t}}{\frac{1}{2}a_j+\omega +\Omega }\ket{\pplusq;-\trion}\bra{\pminusq;+\trion}\nonumber\\
  +& \frac{\ee^{-\Half\gamma t + \ii \epsilon^-_q t}}{\frac{1}{2}a_j+\omega -\Omega }\ket{\pqminus ;\trion+}\bra{\pqplus;\trion-} 
  + \frac{\ee^{-\Half\gamma t + \ii \epsilon^+_q t}}{\frac{1}{2}a_j-\omega +\Omega }\ket{\pqplus ;\trion-}\bra{\pqminus;\trion+}
\label{eqnEvoPertOne10stst}
\end{align}
\end{widetext}
contain the decaying oscillation terms from the ST/TS sector. We have adopted a notation where the nuclear index $\pplus$ ($\pminus$) denotes a nuclear configuration with the $j$th nucleus in the state $\ket{+}$ ($\ket{-}$) while the remaining nuclei can be in an arbitrary state, indicated by $\bar{p}$.
We note furthermore that $\Omega_{pq}+\omega_{pq} = \lar + \qf^p$ and $\Omega_{pq}-\omega_{pq} = \lar + \qf^q$.

The second-order perturbation of the eigenvalues, as first non-trivial correction, is important for the accuracy of the perturbative time evolution. The perturbation to the eigenvalues in second order is
\begin{align}\label{eqnPertEigenvaluesTwoFormal}
  \mu\pe2_\alpha
 &= \bramidket{l\pe0_\alpha}{\Li\pe1}{r\pe1_\alpha}\nonumber\\
 &= \sum_{\beta:\mu\pe0_\beta\not=\mu\pe0_\alpha}
                       \frac{\bramidket{l\pe0_\alpha}{\Li\pe1}{r\pe0_\beta}\bramidket{l\pe0_\beta}{\Li\pe1}{r\pe0_\alpha}}{\mu\pe0_\alpha-\mu\pe0_\beta}.
\end{align} 
Substitution of the components $\alpha=p\pm,q\pm$ yields the corrections to $\omega_{pq}$ and $\Omega_{pq}$,
\begin{equation}\label{eqnPertEigenvaluesTwoExplicit}
\begin{aligned}
 (+\ii\omega_{pq})\pe2 &= \mu\pe2_{p-q-} = \ii(Q^+_p-Q^+_q),\\
 (-\ii\omega_{pq})\pe2 &= \mu\pe2_{p+q+} =-\ii(Q^-_p-Q^-_q),\\
 \ii\Omega_{pq}\pe2 &= \mu\pe2_{p-q+} = \ii(Q^+_p+Q^-_q),\\
-\ii\Omega_{pq}\pe2 &= \mu\pe2_{p+q-} =-\ii(Q^-_p+Q^+_q),
\end{aligned}
\end{equation}
where we define
\begin{equation}\label{eqnPertEigenvaluesTwoExplicitQ}
  Q_p^\pm = \sum_j\frac{a_j^2}{4}\frac{\delta_{p^j,\pm}}{\lar + \qf_p\mp\frac{1}{2}a_j},
\end{equation}
with $\delta_{p^j,+}=\tfrac{1}{2}(1+s^p_j)$, i.e., $1$ if the $j$th nucleus is in the $\ket{+}$ eigenstate and $0$ otherwise, and with $\delta_{p^j,-}=\tfrac{1}{2}(1-s^p_j)$. The denominators $\lar+\qf_p\mp\frac{1}{2}a_j=\lar+\qf_{\bar{p}}$ in \eqn\eqref{eqnPertEigenvaluesTwoExplicitQ} encode the eigenfrequency $\lar+\qf_{\bar{p}}$ of the basis state with the $j$th nucleus taken out. For large $N$, the denominators can be approximated using $\qf_{\bar{p}}\approx\qf_p(N-1)/N$, based on the intuition that the contribution to $\qf_p$ from nucleus $j$ is the average over all nuclei. With this approximation, the denominators are independent on $j$, and can be taken out of the summation, so that we obtain
\begin{align}
 \Omega_{pq}\pe2&= Q^+_p+Q^-_q\nonumber\\
  &= \frac{1}{4}\sum_j a_j^2\left(\frac{\delta_{p^j,+}}{\lar+\frac{N-1}{N}\qf_p} + \frac{\delta_{q^j,-}}{\lar+\frac{N-1}{N}\qf_q}\right).\label{eqnPertEigenvaluesTwoApprox}
\end{align}
The remaining summation is just the sum of the squared couplings, that stands in direct correspondence to the dephasing time. For the Larmor frequencies $\Omega_{pq}$, the second order correction is always positive. In practice, this means the transverse hyperfine coupling leads to an increase of the Larmor frequency. In the particular case of the diagonal part $\Omega_{pp}$, the two deltas in \eqn\eqref{eqnPertEigenvaluesTwoApprox} add up to $1$, so that
\begin{equation}\label{eqnPertEigenvaluesTwoApproxDiagonalAppendix}
 \Omega_{pp}\pe2=\frac{1}{4(\lar+\frac{N-1}{N}\qf_p)}\sum_j a_j^2.
\end{equation}
The diagonal Larmor frequency $\Omega_{pp}=\Omega_{pp}\pe0+\Omega_{pp}\pe2$ (with $\Omega_{pp}\pe0=\lar+\qf_p$) can thus be expressed as a function of $\theta_p$, or equivalently, we could state that the second order shift can be expressed as a function of the zeroth order frequency itself. However, this is only true within the approximation of the couplings $a_j$ being equal to the average. In reality, the values $a_j$ are spread around their average, and consequently the second order shifts are spread around the value given by \eqn\eqref{eqnPertEigenvaluesTwoApproxDiagonalAppendix}. Nevertheless, the deviations are small, given that $a_j\ll\lar$, so that \eqn\eqref{eqnPertEigenvaluesTwoApproxDiagonalAppendix} provides a good estimate.

Expressions for higher order corrections to the frequencies require tedious algebra, but can be estimated to be negligible in view of the following arguments. The corrections to the diagonal matrix elements $\Omega_{pp}$ vanish at odd orders, because a non-vanishing contribution requires an even number of spin flips. For even orders, each increase of the order by $2$ introduces an addition factor $\sum_j a_j^2/\lambda^2$. Although we do not know the coefficients of the latter quantity in the frequency perturbation, its small size in the assumed limit $a_j\ll\lambda$  provides a plausible argument that the corrections of perturbative orders $>2$ are negligible.

\section{Comparison of exact and perturbative solutions for two coupled spins}
\label{appMinimalmodel}
%
%
\begin{figure}[t]
\includegraphics[width=85mm]{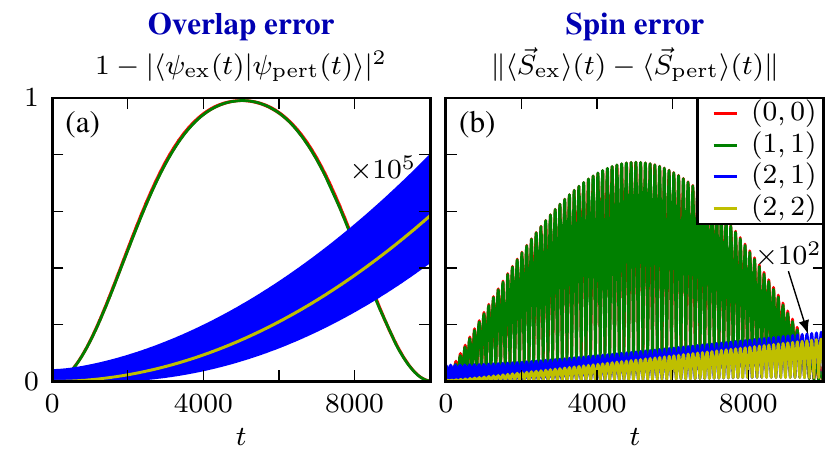}%
\caption{Errors between the exact and perturbative results for Hamiltonian~\eqref{eqnHminimal} with $A=\tilde{A}=0.05$ and some arbitrary initial state $\ket{\psi(0)}$. (a) Overlap errors $1-\abs{\braket{\psi_\mathrm{ex}(t)}{\psi_\mathrm{pert}(t)}}^2$. (b) Spin errors $\norm{\avg{\vec{S}_\mathrm{ex}}(t)-\avg{\vec{S}_\mathrm{pert}}(t)}$. The colors distinguish the perturbation orders, red for $(o_\mathrm{eigenvalues},o_\mathrm{eigenvectors})=(0,0)$, green for $(1,1)$, blue for $(2,1)$, and yellow for $(2,2)$. For the latter two, the error values have been magnified by the factors $10^5$ in (a) and $100$ in (b). The red curve concides with the green one.}
\label{figMinimal}
\end{figure}
%
%
We apply the perturbation theory proposed in Sec.~\ref{secPert} and worked out in Appendix~\ref{appFullPert} to the minimal model of the Hamiltonian dynamics of the central spin model with only one nucleus, in order to answer the question to which order the expansion should be carried out. We write the simplified Hamiltonian of this model as
\begin{equation}\label{eqnHminimal}
  H = \hat{S}^x + A \hat{I}^x\hat{S}^x + \tilde{A}(\hat{I}^y\hat{S}^y+\hat{I}^z\hat{S}^z),
\end{equation}
where we have set the external magnetic field to $1$, and we distinguish $A$ and $\tilde{A}$ as the longitudinal and transverse hyperfine coupling strengths, respectively. The transverse coupling acts as the perturbation parameter, but it is set equal to the longitudinal one at a later stage. In the basis $\{\ket{++},\ket{+-},\ket{-+},\ket{--}\}$, the Hamiltonian is represented by the $4\times4$ matrix
\begin{equation}\label{eqnHminimalmatrix}
  H=\frac{1}{2}
     \begin{pmatrix}
     1+\Half A & 0 & 0 & 0\\
     0 & 1-\Half A & \tilde{A} & 0\\
     0 & \tilde{A} & -1-\Half A & 0\\
     0 & 0 & 0 & -1 + \Half A
    \end{pmatrix}.
\end{equation}
The eigenvalues of this matrix are $(E_1,E_2,E_3,E_4) = (\half+\tfrac{1}{4}A,\half\sqrt{1+\tilde{A}^2}-\tfrac{1}{4}A,-\half\sqrt{1+\tilde{A}^2}+\tfrac{1}{4}A, -\half+\tfrac{1}{4}A)$ and the corresponding eigenvectors are
\begin{gather}
 \ket{r_1} = (1,0,0,0),\quad
 \ket{r_2} = (0,1+\sqrt{1+\tilde{A}^2},\tilde{A},0)/\mathcal{N},\nonumber\\
 \ket{r_3} = (0,-\tilde{A},1+\sqrt{1+\tilde{A}^2},0)/\mathcal{N},\quad
 \ket{r_4} = (0,0,0,1),
\end{gather}
where $\mathcal{N} = [2(1+\tilde{A}^2)+2(1+\tilde{A}^2)^{1/2}]^{1/2}$ is a normalization constant.

We compare this exact result to perturbation theory. The eigenspaces labeled $1$ and $4$ are already exact and therefore the perturbations are trivial. For the other eigenvalues and eigenvectors, we perform an expansion in orders of $\tilde{A}$,
\begin{align}
 E_2 &= (\half-\tfrac{1}{4}A) +\tfrac{1}{4}\tilde{A}^2 +\mathcal{O}(\tilde{A}^3), \nonumber\\
 E_3 &= -(\half-\tfrac{1}{4}A) -\tfrac{1}{4}\tilde{A}^2 +\mathcal{O}(\tilde{A}^3), \nonumber\\
 \ket{r_2} &= \vec{e}_2 +\half \tilde{A} \vec{e}_3 -\tfrac{1}{8}\tilde{A}^2\vec{e}_2 + \mathcal{O}(\tilde{A}^3),\label{eqnHminimalpert}\\
 \ket{r_3} &=\vec{e}_3 -\half \tilde{A} \vec{e}_2 -\tfrac{1}{8}\tilde{A}^2\vec{e}_3 + \mathcal{O}(\tilde{A}^3).\nonumber
\end{align}
In this real and hermitian case, the left eigenvectors are equal to the right eigenvectors. We subsequently derive the time evolution in the exact and in the perturbative case for several orders. The exact time evolution matrix, determined by $\ee^{-\ii t H} = R\ee^{-\ii t D}L$ ($L=R^\dagger$) is
\begin{widetext}
\begin{equation}\label{eqnHminimalevolution}
 \ee^{-\ii t H} =
 \begin{pmatrix}
   \ee^{-\frac{1}{4} \ii A t}\ee^{-\frac{1}{2} \ii t} & 0 & 0 & 0 \\
   0 & \ee^{\frac{1}{4}\ii A t} [\cos \Half rt- \ii r^{-1}\sin \Half r t] & -\ii\tilde{A}r^{-1}\ee^{\frac{1}{4} \ii A t} \sin \Half rt & 0 \\
   0 & -\ii\tilde{A}r^{-1}\ee^{\frac{1}{4} \ii A t} \sin \Half rt & \ee^{\frac{1}{4}\ii A t} [\cos \Half rt+ \ii r^{-1}\sin \Half r t] & 0 \\
 0 & 0 & 0 & \ee^{-\frac{1}{4} \ii A t}\ee^{\frac{1}{2} \ii t}
 \end{pmatrix},
\end{equation}
\end{widetext}
where $r\equiv\sqrt{1+\tilde{A}^2}$. The perturbative result can be found from \eqns\eqref{eqnHminimalpert}, and is equivalent to expansion of each entry in \eqn\eqref{eqnHminimalevolution} into powers of $\tilde{A}$, i.e., $r=1+\tfrac{1}{2}\tilde{A}^2+\ldots$ and $r^{-1}=1-\tfrac{1}{2}\tilde{A}^2+\ldots$.

We compare the exact and perturbative results by examining the errors (i.e., their difference) on the frequencies (energies), and on the coefficients of the diagonal and off-diagonal entries. The frequency (energy) errors are $\mathcal{O}(\tilde{A}^2)$ for the zeroth and first order, and $\mathcal{O}(\tilde{A}^4)$ for second order. The same is true for the diagonal coefficients. The off-diagonal entries are correct up to $\mathcal{O}(\tilde{A})$ for the zeroth order and to $\mathcal{O}(\tilde{A}^3)$ for the first and second order in the eigenvector expansion. In view of the magnitude of $\tilde{A}$ and the time interval we are interested in, we accept errors of quadratic order. Under these conditions, the minimal required perturbation order of the eigenvalues and eigenvectors would be $2$ and $1$, respectively.

This intuition is corroborated by a quantitative analysis of the errors, measured from the overlap between the perturbative and exact wave function and from the difference between the two spin expectation values shown in Fig.~\ref{figMinimal}(a) and (b), respectively. The errors grow rapidly if the perturbation order in the eigenvalues is less than $2$. From Fig.~\ref{figMinimal}(b), we observe that at some moment the spins are almost completely oppositely directed (spin error $\sim 1$). If the eigenvalue perturbation order is chosen equal to $2$, the errors remain smaller over the course of the time interval studied here. (These error values have been magnified in the figure.) If the eigenvalue order is $2$, the accuracy is not increased significantly by including the quadratic order in the  eigenvector. We therefore conclude that perturbation theory of order $2$ in the eigenvalues and order $1$ in the eigenvectors is a reasonable compromise between accuracy and calculation effort.

\section{Truncation of non-diagonal elements of the density matrix}
\label{appTruncation}
As discussed concisely in Sec.~\ref{subsecNumerics}, we ``truncate'' the density matrix by eliminating all matrix elements which have small absolute values. Loosely speaking, the idea behind the truncation is that every spin flip in the time evolution leads to an additional factor of $a_j/\lar$, which are the small perturbation parameters. Thus, many applications of the evolution as described in Appendix~\ref{appFullPert} lead to an exponential increase of nonzero matrix elements in the sparse representation of the density matrix. If we omit this step, the sparse matrix would become increasingly dense, and the calculation intractable.

The precise method of truncation proceeds as follows. At each pulse, all matrix elements $\rho_{pq;\sigma\tau}$ that are non-diagonal in the nuclear degrees of freedom $(p\not=q)$ and smaller than the threshold value $\theta$, i.e., $\abs{\rho_{pq;\sigma\tau}}<\theta$ are set to zero. For the results presented in Sec.~\ref{secModeLocking}, we have used the cutoff value $\theta=4^{-(N+1)}$. Lowering this value leads to a sharp increase in the required computation time, but not to significantly different results. For example, for $N=15$, the relative error between the results at $\theta_1=2^{-40}$ and $\theta_2=2^{-32}$ is of the order of $10^{-6}$ at $T=2000\,\Tpulse$. 

%
\begin{figure}[t]
\includegraphics[width=85mm]{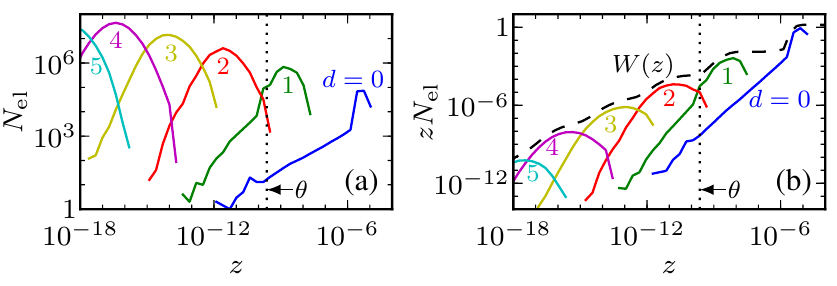}%
\caption{(a) Typical logarithmic histogram of the absolute value $z=\abs{\rho_{pq;\sigma\tau}}$ of the matrix elements of the density matrix. The binning on the horizontal axis is given by $[2^{-(k+1)},2^{-k}]$, $k=0,1,2,\ldots$. The different curves distinguish the distance $d$ to the diagonal (number of different spins between $p$ and $q$). (b) Weight of the matrix elements, defined as number $N_\mathrm{el}$ times value $z$. The dashed curve indicates the accumulated weight $W(z)$ of matrix elements up to $z$, see \eqn\eqref{eqnAccumulatedWeight}. For both plots, $N=15$ and $T^*=1\ns$. The vertical dotted lines show the standard truncation value $\theta=4^{-(N+1)}=2^{-32}$.}
\label{figTruncation}
\end{figure}
%

In Fig.~\ref{figTruncation}, we illustrate the idea behind the approximation: Although  the number $N_\mathrm{el}$ of small matrix elements may be large [see Fig.~\ref{figTruncation}(a)], their weight (number $N_\mathrm{el}$ times value $z$)  is still negligible [see Fig.~\ref{figTruncation}(b)]. For a quantitative estimate, we also explore the \emph{accumulated weight} 
\begin{equation}\label{eqnAccumulatedWeight}
  W(z) = \sum_{\abs{\rho_{pq;\sigma\tau}}<z}\abs{\rho_{pq;\sigma\tau}}
\end{equation}
of all matrix elements smaller than $z$, plotted as the dashed curve in Fig.~\ref{figTruncation}(b). The curve in the plot is an approximation equal to $W(z)-W(\theta_1)$, where $\theta_1$ is a very small cutoff. The precise value of the error $W(\theta_1)$ is unknown, but it is estimated to be small; here $\lesssim 10^{-10}$.
In this case, for $N=15$ with $\theta=2^{-32}$, the neglected accumulated weight is $W(\theta)\sim 10^{-4}$, very small compared to the total weight $W(0)\sim 1$. 

The very weak dependence of the resulting distribution of the longitudinal Overhauser field $O^x$ on the truncation value, is due to the truncated matrix elements values being off-diagonal. Their eventual contribution is roughly their value multiplied by appropriate factors of the small perturbation parameters $a_j/\lambda$. The combination of this observation with the small weights as illustrated by Fig.~\ref{figTruncation}(b) thus explains why truncation of these values has no noticeable effect on the results.

\section{Steady state in zero order}%
\label{appSteadyState}%
In a long time evolution under periodic driving, the system will converge to a steady state. Here, the term steady state refers to periodic time evolution $\rho(t+\Tpulse)=\rho(t)$, for any pair of times separated by one period $\Tpulse$. The time evolution over one period is a combination of the unitary pulse action $\rho\to\tilde{\mathcal{P}}\rho\equiv\mathcal{P}\rho\mathcal{P}^\dagger$ and the evolution between the pulses governed by the Lindblad equation, see Sec.~\ref{subsecEvolutionGeneral}.

In order to gain some basic intuition, we derive the steady state in the zero-order theory, using the explicit time evolution \eqn\eqref{eqnEvoPertZero}. We take the usual action of the $\pi$-pulse given by $\mathcal{P}=\ket{\trion}\bra{\spinup} - \ket{\spinup}\bra{\trion} + \ket{\spindn}\bra{\spindn}$. Without loss of generality, we consider a fixed time in one pulsing period, namely, the moment just before the pulse. The time evolution from one period to the next is then given by
\begin{equation}\label{eqnSteadyStateEvolution}
  \rho((n+1)\Tpulse)
  = \mathcal{U}\rho(n\Tpulse)
  \equiv \ee^{\Tpulse\Li\pe0}\tilde{\mathcal{P}}\rho(n\Tpulse).
\end{equation}
The periodicity condition then defines the steady states as the eigenstates of $\mathcal{U}$ with eigenvalue $1$.

The operator $\mathcal{U}$ can be expressed as a matrix acting on the density matrix  $\rho$ in a vectorized form, similar to the action of the time evolution  $\ee^{\Tpulse\Li\pe0}$ described in Appendix~\ref{appFullPert}. For simplicity, we consider a single nuclear configuration, i.e., we fix the indices $p$ and $q$ in \eqn\eqref{eqnEvoPertZero}, which is justified in view of $\ee^{\Tpulse\Li\pe0}$ and $\tilde{\mathcal{P}}$ being diagonal. For simplicity, we consider the case $p=q$, so that $\omega_{pq}=0$. In addition, we consider the limit $\ee^{-\gamma \Tpulse}\to0$. If we confine ourselves to the SS/TT sector, choosing the same basis order as in Appendix~\ref{appFullPert}, we can write
\begin{widetext}
\begin{equation}
  \mathcal{U}
  =\begin{pmatrix}
    \frac{1}{2} & \frac{1}{2} & 0 & 0 & \frac{1}{2}\\
    \frac{1}{2} & \frac{1}{2} & 0 & 0 & \frac{1}{2}\\
    -\frac{1}{4}\ee^{-\ii \tau}(1+\Gamma^*) & -\frac{1}{4}\ee^{-\ii \tau}(1+\Gamma^*) & \frac{1}{4}\ee^{-\ii \tau}(1-\Gamma^*) & \frac{1}{4}\ee^{-\ii \tau}(1-\Gamma^*) & \frac{1}{2}\ee^{-\ii \tau}\\
    -\frac{1}{4}\ee^{\ii \tau}(1+\Gamma) & -\frac{1}{4}\ee^{\ii \tau}(1+\Gamma) & \frac{1}{4}\ee^{\ii \tau}(1-\Gamma) & \frac{1}{4}\ee^{\ii \tau}(1-\Gamma) & \frac{1}{2}\ee^{\ii \tau}\\
    0 & 0 & 0 & 0 & 0
   \end{pmatrix},
\end{equation}
\end{widetext}
where we define $\tau= \Omega_{pp}\Tpulse$ and $\Gamma= \ii\gamma/(\Omega_{pp}-\ii\gamma)$. This matrix has an eigenvalue equal to $1$ independent of the parameters \footnote{The only other nonzero eigenvalue is $\frac{1}{2}\cos\tau-\frac{1}{2}\abs{\Gamma}\cos(\tau+\phi)$, with $\phi$ given by $\Gamma=\abs{\Gamma}\ee^{\ii\phi}$. For small $\abs{\Gamma}$, this eigenvalue lies close to $\frac{1}{2}$.}. The steady-state density matrix, that is characterized by the eigenstate, is
\begin{equation}\label{eqnSteadyStatePrePulse}
  \bar{\rho} =
  \begin{pmatrix}
    \frac{1}{2} + \bar{s}^x & \bar{s}^z + \ii \bar{s}^y & 0\\
    \bar{s}^z - \ii \bar{s}^y & \frac{1}{2} - \bar{s}^x & 0\\
    0&0&0
  \end{pmatrix},
\end{equation}
with $\bar{s}^x = 0$,
\begin{align}
  \bar{s}^y &= \frac{\sin\tau-\abs{\Gamma}\sin\phi+\abs{\Gamma}\sin(\tau+\phi)}{-4+2\cos\tau-2\abs{\Gamma}\cos(\tau+\phi)},\nonumber\\
  \bar{s}^z &= \frac{\cos\tau+\abs{\Gamma}\cos(\tau+\phi)}{-4+2\cos\tau-2\abs{\Gamma}\cos(\tau+\phi)},\label{eqnSteadyStateSpin}
\end{align}
where we have defined $\phi$ from setting $\Gamma=\abs{\Gamma}\ee^{\ii\phi}$. This density matrix encodes a state for which the spin expectation value is $(\bar{s}^x,\bar{s}^y,\bar{s}^z)$ before each pulse. In the spin-$z$ basis, the pre-pulse steady state is written as $(\frac{1}{2}+\bar{s}^z)\ket{\spinup}\bra{\spinup} + (\frac{1}{2}-\bar{s}^z)\ket{\spindn}\bra{\spindn} + \ii \bar{s}^y\left(\ket{\spindn}\bra{\spinup} - \ket{\spinup}\bra{\spindn}\right)$. From this representation, it is straightforward to determine the post-pulse density matrix as
\begin{equation}\label{eqnSteadyStatePostPulse}
  (\half+\bar{s}^z)\ket{\trion}\bra{\trion} + (\half-\bar{s}^z)\ket{\spindn}\bra{\spindn} + \ii \bar{s}^y\left(\ket{\spindn}\bra{\trion} - \ket{\trion}\bra{\spindn}\right).
\end{equation}
Here, we note that the post-pulse state always points down, if we consider the SS/TT sector only. In particular, the component $\bar{s}^y$ is mapped into the ST/TS sector, which decays to a negligible value over a period of $\Tpulse$. The irrelevance of $\bar{s}^y$ means that the periodicity condition is fulfilled even if $\Tpulse$ is not an integer multiple of the Larmor period, i.e., for any value of $\tau$ (modulo $2\pi$). As a consequence, given the parameters $\gamma$ and $\Tpulse$ as input, no particular value for $\Omega_{pp}$ is singled out as being ``resonant''.

In absence of mode locking, when the frequency distribution is Gaussian, we can assume that the distribution of $\tau$ modulo $2\pi$ is uniform in $[0,2\pi]$. Then the steady state is characterized by the average values of $\bar{s}^y$ and $\bar{s}^z$,
\begin{align}
  \avg{\bar{s}^y}
  &\approx \frac{1}{2\pi}\int_0^{2\pi}\frac{\sin\tau\,\mathrm{d}\tau}{-4+2\cos\tau}
  = 0,\nonumber\\
  \avg{\bar{s}^z}
  &\approx \frac{1}{2\pi}\int_0^{2\pi}\frac{\cos\tau\,\mathrm{d}\tau}{-4+2\cos\tau}
  = \frac{1}{2}-\frac{1}{\sqrt{3}}\approx-0.077,\label{eqnSteadyStateAverageSpin}
\end{align}
in the limit $\abs{\Gamma}\to 0$, i.e., neglecting the effect of the trion decay. The corresponding post-pulse value is $-1/2\sqrt{3}\approx -0.289$. Thus, in the steady state the system acquires a nonzero spin expectation value in the $z$ direction both before and after the pulse.

If the system is maximally mode locked, only a single value of $\tau$ contributes. Considering again the limit $\abs{\Gamma}\to 0$, and assuming a half-integer number of Larmor oscillations in one pulsing period [$\tau\equiv\pi \pmod{2\pi}$] we find that the steady-state pre- and post-pulse expectation values are $\frac{1}{6}$ and $-\frac{1}{6}$, respectively. For an integer number of Larmor oscillations [$\tau\equiv0 \pmod{2\pi}$], both values are equal to $-\frac{1}{2}$. In these two cases the amplitude of the oscillations is not changed by the pulse, but the values are different, and there is a sign flip in the half-integer case that is absent in the integer case.


%

\onecolumngrid
\medskip
\noindent\hrulefill

\bigskip
{\centering\bf\large Erratum\\}

\bigskip

\noindent In previous versions of the article, two figures contain errors:
\begin{itemize}
 \item The values of the transverse components of the Overhauser field, in Figs.~2(e) and 2(f), had been plotted with an incorrect scaling factor on the vertical axis. The correct multiplier on the vertical axis should be $10^{-6}$ instead of $10^{-3}$.
 \item The numbers on the vertical axis in Fig.~3(a), representing the peak growth rate $\eta_t$, did not align properly with the markers. The correct values should run from $-40$ (bottom) to $20$ (top).
\end{itemize}

\noindent The present version includes the corrected figures. Since we have not made quantitative statements based directly on the values represented by the affected figures, the other results and conclusions remain valid without change.
\bigskip

\noindent This Erratum has been published as \href{\doibase 10.1103/PhysRevB.96.199904}{Phys.\ Rev.\ B \textbf{96}, 199904 (2017)}.

\end{document}